\documentclass[12pt,twoside]{article}
\usepackage{amsmath,mathrsfs,bm,amssymb,color}
\newtheorem{theorem}{Theorem}[section]
\newtheorem{proposition}[theorem]{Proposition}
\newtheorem{lemma}[theorem]{Lemma}
\newtheorem{corollary}[theorem]{Corollary}
\newtheorem{definition}[theorem]{Definition}
\newtheorem{example}[theorem]{Example}
\newtheorem{remark}[theorem]{Remark}

\newtheorem{assumption}[theorem]{Assumption}
\newcommand{\BERN}{\ms B_0}
\newcommand{\eq}[1]{\begin{equation}\label{#1}}
\newcommand{\en}{\end{equation}}
\newcommand{\eqn}{\begin{eqnarray*}}
\newcommand{\enn}{\end{eqnarray*}}
\newcommand{\eqnn}{\begin{eqnarray}}
\newcommand{\ennn}{\end{eqnarray}}
\newcommand{\proof}[1][Proof]{{\sc #1.} }
\newcommand{\qed}{\hfill {\bf qed}\par\medskip}
\newcommand{\bi}{\begin{description}}
\newcommand{\ei}{\end{description} }
\newcommand{\bern}{\Psi}
\newcommand{\TTT}{\theta}
\newcommand{\spin}{\TTT}
\renewcommand{\c}{\TTT}

\newcommand{\lkkk }{\left[}
\newcommand{\rkkk}{\right]}
\newcommand{\skima}{\!\frac{}{}\!}
\newcommand{\loc}[1]{L_{\rm loc}^{#1}(\RR^d)}

\newcommand{\VS}{{\bf \s}}
\newcommand{\BM}{\Omega_P}
\newcommand{\PO}{\Omega_N}
\newcommand{\SU}{\Omega_\nu}
\newcommand{\ix}{\int_{\BR}\!\!\!dx}
\newcommand{\bernz}{\bern({\rm p}^2/2)}
\newcommand{\halfp}{\half {\rm p}^2}
\newcommand{\bl}[1]{\begin{lemma}\label{#1}}
\newcommand{\el}{\end{lemma}}
\newcommand{\bc}[1]{\begin{corollary}\label{#1}}
\newcommand{\ec}{\end{corollary}}
\newcommand{\bt}[1]{\begin{theorem}\label{#1}}
\newcommand{\et}{\end{theorem}}
\newcommand{\bp}[1]{\begin{proposition}\label{#1}}
\newcommand{\ep}{\end{proposition}}
\newcommand{\br}[1]{\begin{remark}\label{#1}}
\newcommand{\er}{\end{remark}}
\newcommand{\bd}[1]{\begin{definition}\label{\rm #1}}
\newcommand{\ed}{\end{definition}}

\newcommand{\M}{M_{\mathbb Z_p}}

\newcommand{\CC}{{{\mathbb  C}}}
\newcommand{\U}{{{\rm F}}}
\newcommand{\RR}{{\mathbb  R}}
\newcommand{\BR}{{{\mathbb  R}^d }}
\newcommand{\limn}{\lim_{n\rightarrow\infty}}

\newcommand{\SSS}{{\rm S}}
\newcommand{\kak}[1]{(\ref{#1})}
\newcommand{\LR}{{L^2(\BR)}}
\newcommand{\is}[1]{{\ms E}_{#1}}

\newcommand{\lk}{\left(}
\newcommand{\rk}{\right)}
\newcommand{\lkk}{\left\{}
\newcommand{\rkk}{\right\}}
\newcommand{\ms}[1]{\mathscr{#1}}
\newcommand{\la}{\lambda }
\newcommand{\EE}{\mathbb E}
\newcommand{\ov}[1]{\overline{#1}}

\newcommand{\mmm}[4]
{\left[ \!\!\!\begin{array}{cc}#1&#2\\
#3&#4\end{array}\!\!\!\right]}

\newcommand{\half}{\frac{1}{2}}
\newcommand{\han}{{1/2}}
\newcommand{\hgs}{h_{\mathbb Z_p}}
\newcommand{\HGS}{\hgss}
\newcommand{\hgss}{H_{\mathbb Z_p}^{\bern}}

\newcommand{\Spec}{\mathop{\mathrm{Spec}}\nolimits}

\newcommand{\vvv}[1]
{\lkkk \!\!\!\begin{array}{c}#1\end{array}\!\!\!\rkkk}

\newcommand{\s}{\sigma}

\renewcommand{\d}{\displaystyle}
\newcommand{\non}{\nonumber}

\makeatletter
\@addtoreset{equation}{section}
\makeatother

\title
{\Large \sc Path Integral Representation for  Schr\"odinger Operators with
Bernstein Functions of the Laplacian}
\author{
\small Fumio Hiroshima\\
{\small\it Faculty of Mathematics, Kyushu University}    \\[-0.7ex]
{\small\it  6-10-1 Hakozaki, Fukuoka, 812-8581,  Japan}      \\[-0.7ex]
{\small  {\tt hiroshima@math.kyushu-u.ac.jp}}\\[0.3cm]
\small Takashi Ichinose\\
{\small\it Department of Mathematics,
Kanazawa University}    \\[-0.7ex]
{\small\it Kanazawa, 920-11, Japan}      \\[-0.7ex]
{\small  {\tt ichinose@kenroku.kanazawa-u.ac.jp}  }\\[0.3cm]
\small J\'ozsef L\H{o}rinczi\\
{\it \small School of Mathematics, Loughborough University} \\[-0.7ex]
{\it \small Loughborough LE11 3TU, United Kingdom} \\[-0.7ex]
{\small {\tt  J.Lorinczi@lboro.ac.uk}} \\[-0.7ex]}
\date{\today}

\begin{document}
\maketitle
\setlength{\baselineskip}{18pt}
\begin{abstract}
\noindent
Path integral representations for generalized Schr\"odinger
operators obtained under a class of Bernstein functions of
the Laplacian are established. The one-to-one
correspondence of Bernstein functions with L\'evy
subordinators is used, thereby
the role of Brownian motion entering the standard
Feynman-Kac formula is taken here by subordinated Brownian
motion. As specific examples, fractional and relativistic
Schr\"odinger operators with magnetic field and spin are
covered. Results on self-adjointness of these operators
are obtained under conditions allowing for singular
magnetic fields and singular external potentials as well
as arbitrary
integer and half-integer spin values. This approach also
allows to propose a
notion of generalized Kato class for which
hypercontractivity of the associated
generalized Schr\"odinger semigroup is shown. As a
consequence, dia\-magnetic and energy comparison
inequalities are also derived.

\end{abstract}

\newpage
\section{Introduction}
\subsection{Context and motivation}
Feynman-Kac-type formulae prove to be a useful device in the analysis of spectral properties
of a wide class of self-adjoint operators. Besides their prolific uses in the physics literature,
functional integration poses remarkable new mathematical problems which can be addressed in
terms of modern stochastic analysis.

The Feynman-Kac formula is a functional integral representation of the kernel of the semigroup
generated by the Schr\"odinger operator
\eq{F}
H=\halfp +V,
\en
for which it was originally derived. Here ${\rm p}=-i\nabla$ is the momentum operator and $V$
is a potential. The Laplacian gives rise to an integral representation of the kernel of $e^{-tH}$
in terms of the Wiener measure, while $V$ introduces a density with respect to it. This implies
that the ground state and various other properties of $H$ can be analyzed by running a Brownian
motion under the potential $V$. Standard references on applications to the spectral analysis of
Schr\"odinger operators include \cite{lie73,li80,shi87,sim82}, with updated bibliography in
\cite{sim04}. We also refer to \cite{dd00} for an approach with the Feynman-Kac formula. While
functional integration can be extended to include several other operators also covering quantum
field models (see \cite{LHB09} and references therein), the analysis based on random processes
having almost surely continuous paths remained a basic feature.

In the mathematical physics literature there appear to be relatively few systematic attempts
in going beyond continuous paths to replace them with \emph{c\`adl\`ag} paths (right-continuous
with left limits), also allowing jump discontinuities. On the other hand, such more general
L\'evy processes than Brownian motion prove to be useful in describing important features such as
spin in terms of path measures. Another source of problems leading to paths with jump discontinuities
are models featuring fractional Laplacians.

The aim of the present paper is to construct path integral representations for generalized
Schr\"odinger operators including both non-relativistic and relativistic Schr\"odinger operators
with vector potentials and spin. We propose a thorough study of this problem, extending the methods
developed in \cite{hl08} to the case of L\'evy processes with \emph{c\`adl\`ag} paths.

By a generalized Schr\"odinger operator here we mean a Schr\"odinger operator in which the Laplacian
is replaced by a suitable pseudo-differential operator. Namely, instead of the operator
\eq{1}
\half(\VS  \cdot  ({\rm p}-a))^2+V
\en
studied in \cite{hl08}, where $\VS = (\s_1,\s_2,\s_3)$ are the Pauli matrices and $a$ is
a vector potential, we consider a class of general self-adjoint operators of the form
\eq{2}
\bern \lk \half(\VS  \cdot  ({\rm p}-a))^2 \rk +V,
\en
where $\bern $ is a Bernstein function on the positive semi-axis (see below). In particular,
this class includes not only relativistic Schr\"odinger operators
\eq{3}
\sqrt{(\VS  \cdot ({\rm p}-a))^2+m^2}-m+V
\en
but also more general fractional Schr\"odinger operators
\eq{4}
\left(\frac{1}{2}(\VS  \cdot ({\rm p}-a))^2\right)^{\alpha} + V,
\en
with $\alpha \in (0,1)$. The vector potential plays the role of magnetic field in appropriate
contexts, however, we will use this terminology for all cases we consider, even when they may have
other interpretations.

The application of functional integral techniques to relativistic Schr\"odinger operators, without
magnetic field or spin, has been earlier on addressed in \cite{CMS90}. The process involved is
closely related to $\han$-stable processes, which can be understood in terms of a first hitting
time process of Brownian motion. In the interesting papers \cite{als83, ars91} a path integral for
relativistic Schr\"odinger operators with vector potential and spin $\han$ is presented, however,
in a non-rigorous language.
A functional integral representation also has been established for the Schr\"odinger semigroup
with vector potential in \cite{it86}, applied in \cite{ichi87} and completed in \cite{ichi94},
where, however, the operator concerned was a pseudo-differential operator associated with the
symbol of the classical relativistic Hamiltonian defined through Weyl quantization. It should be
noted that the terms in \kak{2}--\kak{4} involving a vector potential cannot be defined as
pseudo-differential operators associated with simple and plain symbols. A further step has been
made by addressing various problems of potential theory and heat kernel estimates of more general
$\alpha$-stable processes \cite{BB99,BJ07,BKM06,CS97,Ryz02,KS06,GR07}; see also the influential
work \cite{Bak87} involving the Cauchy process. Such processes relate with fractional Schr\"odinger
operators
\eq{frac}
\left(\frac{1}{2}{\rm p}^2\right)^{\alpha} + V
\en
and are motivated by further models of physics, chemistry, biology and, more recently, financial
mathematics \cite{BG90,BBACT02,EK95,MK04}.

Fractional Schr\"odinger operators and stable processes provide just one special case of a
sensible class of extensions. In the present paper we consider generalized Schr\"o\-din\-ger
operators obtained as Bernstein functions of the Laplacian to which we add an external
potential $V$, and in various versions, a vector potential and a contribution from a spin
operator. In a sense, this is the greatest desirable generality as Bernstein functions
with vanishing right limits at the origin stand in a one-to-one correspondence with L\'evy
subordinators. Subordinators are random processes with jump discontinuities and can be
uniquely described by specifying two parameters, the L\'evy measure accounting for the jumps,
and the drift function accounting for the continuous component of the paths. Given a Bernstein
function $\bern$ and a generalized Schr\"odinger operator $H^\bern$ thereby obtained, the
properties of the semigroup $e^{-tH^\bern}$ can now be analyzed
in terms of a subordinated Brownian motion $B_{T^\bern_t}$. Here $T_t^\bern $ is the L\'evy
subordinator uniquely associated with $\bern$. Roughly speaking, $B_{T^\bern _t}$ is a
\emph{c\`adl\`ag} process which samples Brownian paths at random times distributed by the law
of $T^\bern_t$.

\subsection{Main results}
Throughout this paper we will use the following conditions on the vector potential.
\begin{assumption}
\label{fumio4}
The vector potential $a=(a_1,...,a_d)$ is a vector-valued function whose components $a_\mu$,
$\mu=1,...,d$, are real-valued functions. Furthermore, we consider the following regularity
conditions:
\bi
\item[(A1)]
$a\in (\loc 2)^d $.
\item[(A2)]
$a\in (\loc 2)^d $ and $\nabla\cdot a\in \loc 1$.
\item[(A3)]
$a\in (\loc 4)^d$ and $\nabla\cdot  a\in \loc 2$.
\item[(A4)]
$d=3$, $a\in (L_{\rm loc}^4(\RR^3))^3$,
$\nabla\cdot  a
\in L_{\rm loc}^2(\RR^3)$
 and $\nabla\times a\in
 (L_{\rm loc}^2(\RR^3))^3$.
\ei
\end{assumption}

Since we discuss several variants of Schr\"odinger operators, different by whether they do
or do not include spin, it is appropriate to explain here the notation. We define
the spinless operator through a quadratic form for $a$ satisfying (A1) and denote it by
\eq{sss5}
h=\half({\rm p}-a)^2 \qquad (\mbox{with no spin}).
\en
A Schr\"odinger operator with spin $\han$ also is defined through a quadratic form and will
be denoted by
\eq{cons}
h_\han=\half(\s\cdot({\rm p}-a))^2\qquad (\mbox{with spin}).
\en
Using a suitable unitary map, we transform $h_\han$ on the space $L^2(\RR^3;\mathbb C^2)=
L^2(\RR^3)\otimes\CC^2$ to a self-adjoint operator $h_{\mathbb Z_2}$ on $L^2(\RR^3\times
\mathbb Z_2)$. Here $\mathbb Z_2=\{-1,1\}$ describes the state space of a two-valued spin
variable. Furthermore, we generalize spin from $\mathbb Z_2$ to $\mathbb Z_p$ and denote a
so obtained Schr\"odinger operator by
\eq{sss4}
\hgs \qquad (\mbox{with generalized spin})
\en
acting on $L^2(\BR\times\mathbb Z_p)$, for $d\geq 1$ and $p\geq 2$. The relativistic versions
of \kak{sss5} and \kak{cons} will be denoted by
\eq{relv}
\begin{array}
{l}
h^{\rm rel}=
\sqrt{({\rm p}-a)^2+m^2}-m, \quad m \geq 0 \\ \ \\
h_\han^{\rm rel}=   \sqrt{(\s\cdot ({\rm p}-a))^2+m^2}-m, \quad m \geq 0.
\end{array}
\en

In this paper we will consider generalized versions of \kak{relv}.
%
Let $\bern$ be a Bernstein function. Our main objects are
\eq{ham}
\begin{array}{ll}
H^\bern  =\bern(h)+V & (\mbox{with no spin}),\\
&
\\
\HGS=\bern(\hgs)+V & (\mbox{with generalized spin}).
\end{array}
\en
In particular, $$\bern(u)=\sqrt{2u+m^2}-m$$ corresponds to \kak{relv}. Under Assumptions (A2)
(resp. (A3)), we will show that $C_0^\infty(\BR)$ is a form core (resp. operator core) of both
$\bern(h)$ and $\bern(\hgs)$. This is the content of Theorems \ref{main2} and \ref{yy4} below.

The key results of this paper are the functional integral representations of $e^{-tH^\bern}$
and $e^{-t\HGS}$ derived under Assumption (A2) for bounded potentials $V$. They are presented
in Theorems \ref{3t} and \ref{MAIN}, respectively. These are then further generalized to more
singular potentials in Theorems \ref{f} and \ref{ko39}. Recall that the standard
Feynman-Kac-It\^o formula says that
\eq{fki1}
(f, e^{-t(h+V)}g)=\ix\EE_P^x\left[\ov{f(B_0)}g(B_t) e^{-i\int_0^t a(B_s)\circ dB_s}
e^{-\int_0^t V(B_s) ds}\right],
\en
with $d$-dimensional Brownian motion $(B_t)_{t\geq0}$ on Wiener space $(\Omega_P,\ms F_P,P^x)$,
where the stochastic integral in the exponent is Stratonovich integral. For $H^\bern=\bern(h)+V$
this formula modifies to (see Theorem \ref{f} below)
\eq{fki2}
(f, e^{-t(\bern(h)+V)}g)=\ix\EE_{P\times\nu}^{x,0} \left[\ov{f(B_0)}g(B_{T_t^\bern})
e^{-i\int_0^{T_t^\bern} a(B_s)\circ dB_s} e^{-\int_0^t V(B_{T_s^\bern}) ds}\right],
\en
where $T_t^\bern$ is the L\'evy subordinator on a probability space $(\Omega_\nu,\ms F_\nu,\nu)$
associated with $\bern$. In particular, it should be noted that the integrands change as
$$
\exp\lk -i\int_0^{t} a(B_s)\circ dB_s \rk \dashrightarrow
\exp\lk -i\int_0^{T_t^\bern} a(B_s)\circ dB_s \rk
$$
and
$$
\exp\lk -\int_0^t V(B_s) ds\rk \dashrightarrow \exp\lk -\int_0^t V(B_{T_s^\bern}) ds\rk.
$$
A similar situation occurs in the case including a generalized spin, see Theorem \ref{ko39} below.
By means of these formulae we are able to extend the definition of generalized Schr\"odinger
operators $H^\bern$ and $\HGS$ to the case of external potentials having singularities.

Having the functional integral representations at hand allows us to construct a strongly continuous
symmetric Feynman-Kac semigroup for a large class of potentials $V$ which we call $\bern$-Kato class.
This will be dealt with in Theorem \ref{k5}. The generator of this semigroup can be identified as a
self-adjoint operator, which we denote by
\eq{ss1}
K^\bern   \quad (\mbox{with $\bern$-Kato class potential}).
\en
This offers then a notion of generalized Schr\"odinger operator with vector potential for $\bern$-Kato
potentials. As a further result, we show hypercontractivity of the semigroup $e^{-tK^\bern }$ in Theorem
\ref{hyp}.

As corollaries of Theorems \ref{3t} and \ref{MAIN}, by choosing $\bern(u)=\sqrt{2u+m^2}-m$ mentioned
above we obtain the functional integral representations of the relativistic Schr\"odinger operators
\eq{sss3}
\begin{array}
{ll}
h^{\rm rel}+V=  \sqrt{({\rm p}-a)^2+m^2}-m+V & (\mbox{with no spin}),\\ & \\
h_\han^{\rm rel}+V=   \sqrt{(\s\cdot ({\rm p}-a))^2+m^2}-m+V & (\mbox{with spin $\han$})
\end{array}
\en
in Theorems \ref{ko50} and \ref{ko52}, respectively, and derive energy comparison inequalities.
Our results improve and generalize those of \cite{BHL00, CMS90, it86, als83, ars91,sim82,GV81}.
Further applications to relativistic quantum field theory are discussed in \cite{hir09,hs09,L09a,L09b}.

The paper is organized as follows. In Section 2 we  discuss the details of the relationship between
Bernstein functions $\bern$ and L\'evy subordinators $(T_t^\bern)_{t\geq 0}$. In Section 3 we consider
the spinless case. We establish the functional integral representation for their semigroup and obtain
diamagnetic inequalities. Furthermore, we show essential self-adjointness of $\bern(h)$ on
$C_0^\infty(\BR)$. In Section 4, we define the space of $\bern$-Kato class potentials and discuss their
relationship with the L\'evy measure of the associated subordinators. Also, we prove hypercontractivity
of the generalized Schr\"odinger semigroups obtained for this class. In Section 5 we consider generalized
Schr\"odinger operators with spin. We extend $\pm1$ spins to spins of $p$ possible orientations by
describing them in terms of the cyclic group of the $p$th roots of unity. This gives rise to a random
process driven by a weighted sum of $p$ independent Poisson variables of intensity 1. As a corollary,
we derive diamagnetic inequalities. Finally, in Section 6 we give a functional integral representation
of the relativistic Schr\"odinger operator with and without spin as a special case.

\section{Bernstein functions and L\'evy subordinators}
We start by considering some basic facts on Bernstein functions and their connection with
subordinators. For standard definitions and results on Bernstein functions we refer to
\cite{B55,bf73}, for L\'evy processes to \cite{Sato99}, to \cite{Ber99} for a detailed study
on L\'evy subordinators, and to \cite{H69,SV09} for details on subordinated Brownian motion.

Bernstein functions appear in the analysis of convolution semigroups, in particular they are
a key concept in Bochner's theory of subordination.
\begin{definition}[\textbf{Bernstein function}]
\label{44}
{\rm
Let
$$
\ms B=\left\{ f\in C^\infty((0,\infty))\,
\left|\, f(x) \geq 0 \;\; \mbox{and} \;\; (-1)^n \lk \frac{d^nf}{dx^n}\rk(x) \leq 0 \;\;
\mbox{\rm for all}\;\; n=1,2,...,\right.\right\}.
$$
An element of $\ms B$ is called a \emph{Bernstein function}. We also define the subclass
$$
\BERN =\lkk f\in \ms B \left|\lim_{u\rightarrow 0+}f(u)=0\right.\rkk.
$$
}
\end{definition}

Bernstein functions are positive, increasing and concave.
$\ms B$ is a convex cone containing
the nonnegative constants. Examples of functions in $\BERN $ include $\bern(u)=cu^\alpha$, $c\geq 0$,
$0<\alpha\leq 1$,
and $\bern(u)=1-e^{-au}$, $a\geq0$.

A real-valued function $f$ on $(0,\infty)$ is a Bernstein function if and only if $g_t:=e^{-tf}$ is
a completely monotone function for all $t > 0$, i.e., exactly when $(-1)^n \frac{d^n g_t}{dx^n} \geq
0$, for all integers $n \geq 0$. On the other hand, a result by Bernstein says that a function is
completely monotone if and only if it is the Laplace transform of a positive measure, which for each
such function is unique. This leads to the following integral representation of Bernstein functions.

\begin{definition}[\textbf{Class $\ms L$}]
\label{5}
{\rm
Let $\ms L$ be the set of Borel measures $\la$ on $\RR\setminus\{0\}$ such that
\begin{itemize}
\item[(1)]
$\la((-\infty,0))=0$;
\item[(2)]
$\d \int_{\RR\setminus\{0\}} (y\wedge1)\la (dy)<\infty$.
\end{itemize}
\label{classL}
}
\end{definition}
Note that each $\la\in \ms L$ satisfies that $\int_{\RR\setminus\{0\}}(y^2\wedge 1) \la(dy)<\infty$ so that
$\la$ is a L\'evy measure.

Denote $\RR_+=[0,\infty)$. We give the integral representation of Bernstein functions with vanishing
right limits at the origin.
\begin{proposition}
For every Bernstein function $\bern\in \BERN $ there exists $(b,\la)\in \RR_+ \times \ms L$ such that
\eq{6}
\bern(u)=bu+\int_0^\infty (1-e^{-uy})\la(dy).
\en
Conversely, the right hand side of \kak{6} is in $\BERN $ for each pair $(b,\la)\in \RR_+\times \ms L$.
\end{proposition}
For a given $\bern\in \BERN$, the constant $b$ is uniquely determined by $b=\lim_{u\rightarrow \infty}
\bern (u)/u$. Moreover, since $\frac{d\bern}{du}=b+\int_0^\infty ye^{-yu}\la(dy)$ and $\frac{d\bern}{du}$
is a completely monotone function, the measure $\la$ is also uniquely determined; for details, see
\cite[Theorem 9.8]{bf73}. Thus the map $\BERN \rightarrow \RR_+\times\ms L$, $\bern\mapsto(b,\la)$ is a
one-to-one correspondence.

Next we consider a probability space $(\SU, \ms F_\nu, \nu)$ given and the following special class of
L\'evy processes.
\begin{definition}[\textbf{L\'evy subordinator}]
\label{7}
{\rm
A random process $(T_t)_{t\geq 0}$ on $(\SU, \ms F_\nu, \nu)$ is called a \emph{(L\'evy) subordinator}
whenever
\begin{itemize}
\item[(1)]
$(T_t)_{t\geq0}$ is a L\'evy process starting at 0, i.e., $\nu(T_0=0)=1$;
\item[(2)]
$T_t$ is almost surely non-decreasing in $t$.
\end{itemize}
}
\end{definition}
Subordinators have thus independent and stationary increments, almost surely no negative jumps, and
are of bounded variation. These properties also imply that they are Markov processes.

{}

Let $\ms S$ denote the set of subordinators on $(\SU, \ms F_\nu, \nu)$.
In what follows we denote expectation by $\mathbb E_m^x[\cdots]=\int \cdots dm^x$ with respect to
the path measure $m^x$ of a process starting at $x$.
\bp{9}
Let $\bern\in \BERN $ or, equivalently, a pair $(b,\la)\in \RR_+\times \ms L$ be given.
Then there exists a unique $(T_t )_{t\geq 0}\in \ms S$ such that
\eq{10}
\EE_{\nu}^0[e^{-u T_t}]=e^{-t \bern(u)}.
\en
Conversely, let $(T_t)_{t\geq0}\in \ms S$. Then there exists $\bern\in \BERN $, i.e., a
pair $(b,\la)\in \RR_+\times \ms L$ such that \kak{10} is satisfied.
\ep
In particular,
(\ref{6}) coincides with the L\'evy-Khintchine formula for Laplace exponents of subordinators.

By the above there is a one-to-one correspondence between $\BERN $ and $\ms S$, or equivalently,
between $\BERN$ and $\RR_+\times\ms L$. For clarity, we will use the notation $T_t^\bern$ for the
L\'evy subordinator associated with $\bern \in \BERN $.

\begin{example}[\textbf{Stable processes}]
\rm{
Let $b=0$, $0<\alpha<1$  and $\la\in\ms L$ be defined by
$$
\la(dy)=\frac{\alpha}{\Gamma(1-\alpha)} \frac{1_{(0,\infty)}(y)}{y^{1+\alpha}} dy,
$$
where $\Gamma$ denotes the Gamma function. Then $\bern(u)=u^\alpha\in\BERN $ and the
corresponding subordinator $T_t^\bern$ is given by
$$
\EE_\nu^0[e^{-uT_t^\bern}]=e^{-tu^\alpha}.
$$
}
\end{example}

\begin{example}[\textbf{First hitting time}]
\rm{
Since $\bern(u)=\sqrt{2u+m^2}-m  \in\BERN $ for $m\geq0$, there exists $T_t^\bern \in\ms S$
such that
$$
\EE_\nu^0[e^{-u T_t^\bern}] = \exp\lk -t(\sqrt{2u+m^2}-m) \rk.
$$
This case is thus related to
 the one-dimensional $\han$-stable process and it is known that
the corresponding subordinator $T_t^\bern$ can be represented as the first hitting time process
\eq{bo4}
T_t^\bern=\inf\{s>0 \,|\, B_s+m s = t\}
\en
for one-dimensional Brownian motion $(B_t)_{t\geq 0}$. In this case, moreover, the
distribution also is known exactly to be
\eq{enta}
\rho(r,t)=\frac{t}{\sqrt{2\pi r^3}} e^{m t} \exp\lk -\half\lk \frac{t^2}{r^2}+m^2 r \rk \rk.
\en
}
\end{example}

\begin{example}[\textbf{Hyperbolic L\'evy motion}]
\rm{
A specific case studied in mathematical finance \cite{EK95} is
$$
\bern(u) = -\log \left(\frac{a
K_1(\sqrt{a^2+b^2u^2})}
{K_1(a)\sqrt{a^2+b^2u^2}} \right), \quad a,b > 0,
$$
where $K_1$ is the modified Bessel function of the third kind with index 1. This is a purely
discontinuous process with L\'evy measure
$$
\la _{a,b}(dy) = \left(\frac{1}{\pi^2y}\int_0^\infty
\frac{e^{-y\sqrt{2x+(a/b)^2}}} {J_1^2(b\sqrt{2x})+Y_1^2(b\sqrt{2x})}
\frac{dx}{x} + \frac{e^{-y}}{y}\right)1_{(0,\infty)}(y) dy,
$$
where $J_1$ and $Y_1$ are the Bessel functions of the first and second kind, with index 1.
}
\end{example}

\section {Spinless case}
\subsection{Generalized Schr\"odinger operators with no spin}
Now we define the class of generalized Schr\"odinger operators on $L^2(\BR)$,  which we consider
in this paper. In order to cover interactions with a magnetic field we add a vector potential
to the momentum operator. Let $\partial_{x_\mu}:\ms D'(\BR)\rightarrow \ms D'(\BR)$, $\mu=1,...,d,$
denote the $\mu$th  derivative on the Schwartz distribution space $\ms D'(\BR)$. With the notation
$
{\rm p} = -i \nabla$
and $
\nabla=(\partial_{x_1},\ldots,\partial_{x_d})$,
the Schr\"odinger operator with vector potential $a$ is formally given by $\half({\rm p}-a)^2$.
We will define it as a self-adjoint operator rigorously through a quadratic form.

Let ${\rm D}_\mu={\rm p}_\mu-a_\mu$, $\mu=1,...,d$. Define the quadratic form
\eq{y1}
q(f,g)=\sum_{\mu=1}^d({\rm D}_\mu f, {\rm D}_\mu g)
\en
with domain
\eq{y2}
Q(q)=\left\{f\in \LR\,|\, {\rm D}_\mu f\in \LR,\, \mu=1,...,d\right\}.
\en
It can be seen that $Q(q)$ is complete with respect to the  norm $\|f\|_q=\sqrt{q(f,f)+\|f\|^2}$ under
Assumption (A1). Thus $q$ is a non-negative closed form and there exists a unique self-adjoint operator
$h$ satisfying
\eq{y123}
(h f,g)=q(f,g), \quad f\in D(h), \quad g\in Q(q),
\en
with domain
\eq{y1234}
D(h)=\left\{f\in Q(q)\,|\,q(f,\cdot)\in \LR'\right\}.
\en
The self-adjoint operator $h$ is our main object in this section. We summarize some facts about the form
core and operator core of $h$ \cite{sim79, ls}.

\bp{y4}
(1) Let Assumption (A1) hold. Then $C_0^\infty(\BR)$ is a form core of $h$.
(2) Let Assumption (A3) hold. Then $C_0^\infty(\BR)$ is an operator core for $h$.
\ep
Note that in case (2) of Proposition \ref{y4},
$$
hf=\half {\rm p}^2  f -a\cdot {\rm p} f+\lk -\half a\cdot a-({\rm p}\cdot a)\rk f.
$$

\begin{definition}\rm{(\textbf{Generalized Schr\"odinger operator with vector potential
and bounded $V$})}
{\rm
Let $\bern\in \BERN $ and take Assumption (A1). Whenever $V$ is bounded we call
\eq{bo3}
H^\bern=\bern(h)+V
\en
\emph{generalized Schr\"odinger operator with vector potential} $a$.
}
\end{definition}
Note that $\bern\geq0$ and $\bern(h)$ is defined through the spectral projection of the
self-adjoint operator $h$. Furthermore, $H^\bern  $ is self-adjoint on the domain $D(\bern(h))$
as $V$ is bounded.

\subsection{Essential self-adjointness}
\bt{main2}
Take $\bern\in\BERN$.
\begin{itemize}
\item[(1)]
Let Assumption (A3) hold. Then $C_0^\infty(\BR)$ is an operator core of $\bern (h)$.
\item[(2)]
Let Assumption (A1) hold. Then $C_0^\infty(\BR)$ is a form core of $\bern (h)$.
\end{itemize}
\et
\proof
(1) Recall the representation (\ref{6}).
Since  we have $\int_0^1 y
\la(dy)<~\infty$ and $\int_1^\infty \la(dy)< \infty$ by Definition  \ref{classL},
there exist non-negative constants $c_1$
and $c_2$ such that $\bern(u)\leq c_1u +c_2$ for all $u\geq 0$. This gives the bound
\eq{f2}
\|\bern(h) f\|\leq
 c_1\|h f\|+ c_2\|f\|
\en
for all $f\in D(h)$. Hence $C_0^{\infty}({\mathbb  R}^d)$ is contained in $D(\bern(h))$. Since
$\bern(h)$ is a non-negative self-adjoint operator, $\bern(h)+1$ has a bounded inverse, and we
use that $C_0^{\infty}({\mathbb  R}^d)$ is a core of $\bern(h)$ if and only if $\bern(h)
C_0^{\infty}({\mathbb  R}^d)$ is dense in $L^2({\mathbb  R}^d)$.
Let $g \in L^2({\mathbb  R}^d)$ and suppose that $(g, (\bern(h)+1)f) =0$, for all $f \in C_0^{\infty}
({\mathbb  R}^d)$. Then $C_0^\infty(\BR)\ni f\mapsto (g, \bern(h) f)=-(g,f)$ defines a continuous
functional which can be extended to $\LR$. Thus $g \in D(\bern(h))$ and $0= ((\bern(h)+1)g, f)$.
Since $C_0^{\infty} ({\mathbb  R}^d)$ is dense, we have $(\bern(h)+1)g =0$, and hence $g=0$ since
$\bern(h)+1$ is one-to-one, proving the assertion.

(2) Note that $\|\bern(h)^{1/2}f\|^2 \leq c_1\|h^{1/2}f\|^2 + c_2\|f\|^2$ for $f \in Q(h)=D(h^{1/2})$,
and $C_0^{\infty}({\mathbb  R}^d)$ is contained in $Q(\bern(h))= D(\bern(h)^{1/2})$. Since
$\bern(h)^{1/2}+1$ has also bounded inverse, it is seen by the same argument as
above that $C_0^{\infty}({\mathbb  R}^d)$ is a core of $\bern(h)^{1/2}$ or a form core of $\bern(h)$.
\qed

\subsection{Singular magnetic fields}
Before constructing  a functional integral representation of $e^{-th}$, we extend stochastic
integration to a class including $\loc 2$ functions since the vector potentials we consider may
be more singular.

Let $(B_t)_{t\geq 0}$ denote $d$-dimensional Brownian motion starting at $x\in\BR$ on standard
Wiener space $(\BM,{\ms F_P}, P^x)$. Let $f$ be a $\CC^d$-valued Borel measurable function on $\BR$
such that
\eq{y16}
\mathbb E_P^x\lkkk \int_0^t |f(B_s)|^2 ds \rkkk <\infty.
\en
Then the stochastic integral $\int_0^t f(B_s) \cdot dB_s$ is defined as a martingale and the It\^o
isometry
$$
\mathbb E_P^x\left[\left|\int_0^t f(B_s)\cdot  dB_s\right|^2\right] =
\mathbb E_P^x\left[\int_0^t |f(B_s)|^2 ds \right]
$$
holds. However, vector potentials $a$ under (A.1) of Assumption \ref{fumio4} do not necessarily satisfy
\kak{y16}. As we show next, a stochastic integral can indeed be defined for a wider class of
functions than \kak{y16}, and then $\int_0^t f(B_s) \cdot dB_s$ will be defined as a local martingale
instead of a martingale. This extension will allow us to derive a functional integral representation
of $e^{-th}$ with $a\in\lk \loc 2\rk^d $.

Consider the following class 
of vector valued functions on $\BR$.
\begin{definition}
{\rm
We say that $f=(f_1,...,f_d) \in \ms E_{\rm loc}$ if and only if for all $t \geq 0$
\eq{y7}
P^x \lk \int_0^t |f(B_s)|^2 ds < \infty \rk = 1.
\en
}
\end{definition}
Let $R_n(\omega)=n \wedge \inf \lkk t\geq 0 \left|\int_0^t |f(B_s(\omega))|^2 ds\geq n \right.\rkk$
be a sequence of stopping times with respect to the natural filtration $\ms F^P_t= \s(B_s, 0\leq s\leq t)$.
Define
\eq{y19}
f_n(s,\omega)=f(B_s(\omega))1_{\{R_n(\omega)>s\}}.
\en
Each of these functions satisfies $\int_0^\infty |f_n(s,\omega)|^2 ds=\int_0^{R_n} |f_n(s,\omega)|^2 ds
\leq n$. In particular, we have $\mathbb E_P^x\left[\int_0^t |f_n|^2 ds\right]<\infty$ and thus
$\int_0^t f_n\cdot  dB_s$ is well defined. Moreover, it can be seen that
\eq{y199}
\int_0^{t\wedge R_m} f_n(s,\omega) \cdot dB_s=\int_0^t f_m (s,\omega)\cdot dB_s
\en
for $m<n$.
\begin{definition}
{\rm
For $f\in \ms E_{\rm loc}$ we define the integral
\eq{y6}
\int_0^t f(B_s) \cdot dB_s := \int_0^t  f_n(s,\omega) \cdot  dB_s, \quad 0\leq t\leq R_n.
\en
}
\end{definition}
This definition is consistent with \kak{y199}.

\bl{y17}
$\ms E_{\rm loc}$ has properties below:
\begin{itemize}
\item[(1)]
Let $f\in \ms E_{\rm loc}$. Suppose that a sequence of step functions $f_n$, $n=1,2,...$,
satisfies $\int_0^t|f_n(B_s)-f(B_s)|^2 ds\rightarrow 0$ in probability as $n\rightarrow
\infty$. Then
$$
\lim_{n\rightarrow \infty}\int_0^t f_n(B_s)\cdot  dB_s= \int_0^t f(B_s) \cdot dB_s
\quad \mbox{in probability}.
$$
\item[(2)]
$(\loc 2)^d \subset \ms E_{\rm loc}$.
\item[(3)]
Let $a\in \lk \loc 2\rk^d $ and $\nabla\cdot  a\in \loc 1$. Then
$$
\left|\int_0^t a(B_s)\cdot  dB_s +\half \int_0^t \nabla\cdot  a (B_s) ds\right|<\infty
\quad\mbox{almost surely.}
$$
\end{itemize}
\el
\proof
(1) is standard. To see (2)
take $f\in \lk \loc 2\rk^d $,
then
$$
\mathbb E_P^x\left[\int_0^t \chi_\xi (B_s) |f(B_s)|^2 ds\right]<\infty,\quad \xi>0,
$$
for any indicator function $\chi_\xi$ of the set $\prod_{\mu=1}^d [-\xi,\xi]$.
Hence $\int_0^t
\chi_\xi(B_s) |f(B_s)|^2  ds<\infty$ for almost all $\omega$. For each $\omega$
there exists $b(\omega)$ such that $\sup_{0\leq s\leq t} |B_s(\omega)|<b(\omega)$.
Take $\xi =\xi(\omega)$ such that $\xi>b(\omega)$. Then $\int_0^t  |f(B_s(\omega))|^2  ds =
\int_0^t \chi_\xi(B_s(\omega)) |f(B_s(\omega))|^2 ds <\infty$, implying
$P^x \lk \int_0^t |f(B_s)|^2  ds <\infty \rk=1$, thus (2) follows. To see (3), note that
$$
\mathbb E_P^x \left[\left|\int_0^t \chi_\xi(B_s) \nabla\cdot a(B_s) ds \right| \right]
\leq
\int_0^t ds \int_\BR  dy \chi_\xi(sy) |(\nabla \cdot a)(sy)| \frac{t e^{-|y|^2/2}}{(2\pi)^{d/2}}
<\infty
$$
for any indicator function $\chi_\xi$, whence follows that $\left|\int_0^t \nabla\cdot
a(B_s)ds \right|<\infty$ for almost every $\omega$. Thus (3) is obtained.
\qed
For $a\in (\loc 2)^d$ such that $\nabla\cdot a\in \loc 1$, we denote
$$
\int_0^t a(B_s) \circ dB_s=\int_0^t a(B_s)\cdot dB_s +\half \int_0^t\nabla\cdot a(B_s) ds.
$$

\bp{y77}
Under Assumption (A2) we have
\eq{y8}
(f,e^{-th }g)=\int_\BR dx \mathbb E_P^x\left[\overline{f(B_0)} g(B_t) e^{-i\int_0^t a(B_s)
\circ dB_s}\right].
\en
\ep
\proof
Equality \kak{y8} is well known as the Feynman-Kac-It\^o formula, which in \cite[Theorem 15.5]{sim04}
was shown for $a\in L_{\rm loc}^2(\BR)$, however, with $\nabla\cdot a=0$. We provide a proof of \kak{y8}
under Assumption (A2) for a self-contained presentation.

By using a mollifier we can take a sequence $a_n \in (C_0^\infty(\BR))^d$, $n=1,2,...$, such that
$a_n\rightarrow a$ in $(L_{\rm loc}^2)^d$ and $\nabla\cdot a_n \rightarrow \nabla\cdot a$ in $L^1_{\rm loc}$
as $n\rightarrow\infty$. Let $\chi_R=\chi(x^1/R)\cdots \chi(x^d/R)$, $R\in\mathbb N$, where $\chi\in
C_0^\infty(\RR)$ such that $0 \leq \chi\leq 1$, $\chi(x)=1$ for $|x|<1$ and $\chi(x)=0$ for $|x|\geq 2$.
Denote $h = h(a)$.
Since
$\chi_R a_n\rightarrow \chi_R a$ as $n\rightarrow \infty$ in $(L_{\rm loc}^2)^d$ and
$\chi_R a\rightarrow a$ as $R\rightarrow \infty$ in $(L_{\rm loc}^2)^d$,
it follows \cite[Lemma 5 (3.17)]{ls}
that $e^{-th (\chi_R a_n )}\rightarrow e^{-th (\chi_R a)}$ as $n\rightarrow\infty$ and $e^{-th (\chi_R a)}
\rightarrow e^{-th (a)}$ as $R\rightarrow\infty$ in strong sense. Furthermore, \kak{y8} remains true for
$a$ replaced by $\chi_R a_n \in (C_0^\infty(\BR))^d$.

Since $\chi_Ra_n \in (C_0^\infty(\BR))^d$ and $\chi_R  a_n \rightarrow \chi_R a$ in $(L^2)^d$ as
$n\rightarrow\infty$, it follows that
\eq{y9}
\int_0^t \chi_R  (B_s) a_n (B_s) \cdot dB_s
\rightarrow
\int_0^t \chi_R  (B_s) a(B_s) \cdot dB_s
\en
almost surely and since $\nabla\cdot(\chi_R a_n)=(\nabla\chi_R)\cdot a_n+\chi_R(\nabla\cdot a_n)
\rightarrow (\nabla\chi_R)\cdot a+\chi_R(\nabla\cdot a)$ in $L^1(\BR)$, it furthermore follows that
\eq{y12}
\int_0^t \nabla \cdot (\chi_R  (B_s) a_n (B_s)) ds \rightarrow \int_0^t (\nabla \chi_R (B_s))
\cdot a (B_s)ds+\chi_R  (B_s) (\nabla\cdot a(B_s)) ds
\en
strongly in $L^1(\BM,dP^x)$. Thus there exists  a subsequence $n'$ such that \kak{y9} and
\kak{y12} with $n$ replaced by $n'$ hold almost surely. Hence \kak{y8} results by a limiting
argument for $a$ replaced by $\chi_R a$.
Let
\begin{eqnarray*}
&&
\Omega_+(R)=\{\omega\in\BM \,|
\max_{0\leq s\leq t,1\leq\mu\leq d} B_s^\mu(\omega)\leq R\},\\
&&
\Omega_-(R)=\{\omega\in\BM \,|
\min_{0\leq s\leq t,1\leq\mu\leq d} B_s^\mu(\omega)\geq-R\}
\end{eqnarray*}
and
$$
I(R)=\left|\int_0^t \chi_R (B_s) a(B_s)\cdot dB_s-\int_0^ta(B_s)\cdot dB_s\right|.
$$
We show that $I(R)\rightarrow 0$ in probability as $R\rightarrow\infty$. Note that the random
variables
$\max_{0\leq s\leq t} B_s^\mu(\omega)$ and $\min_{0\leq s\leq t} B_s^\mu(\omega)$ have the same
distribution and
$$
P(\Omega_-(R))=P(\Omega_+(R))=\prod_{\mu=1}^d
P(|B_t^\mu|\leq R) = \lk \frac{2}{\sqrt{2\pi t}} \int_0^R e^{-y^2/(2t)} dy \rk^{d}.
$$
Since $\chi_R(B_s)=1$ for all $0\leq s\leq t$ on $\Omega_+(R)\cap \Omega_-(R)$, $I(R)=0$ on
$\Omega_+(R)\cap \Omega_-(R)$, we have
\begin{eqnarray*}
P\left(I(R)\geq \epsilon\right) = P(I(R)\geq \epsilon,\, \Omega_+(R)^c\cup \Omega_-(R)^c)
\leq
2 \lk \frac{2}{\sqrt{2\pi t}} \int_R^\infty e^{-y^2/(2t)} dy \rk^{d}.
\end{eqnarray*}
Hence $\lim_{R\rightarrow\infty} P(I(R)\geq \epsilon)=0$. Thus there exists a subsequence $R'$ such
that $\int_0^t \chi_{R'} (B_s) a(B_s)\cdot dB_s \rightarrow \int_0^t  a(B_s)\cdot dB_s$ almost surely
as $R'\rightarrow \infty$. In a similar way it is seen that $\int_0^t \chi_{R''} (B_s) \nabla \cdot
a(B_s)ds \rightarrow \int_0^t  \nabla\cdot a(B_s)ds$ as $R''\rightarrow \infty$ almost surely for
some subsequence $R''$ of $R'$. Moreover,
\eq{y14}
\int_0^t \nabla \chi_R (B_s) \cdot a(B_s)ds = \frac{1}{R} \int_0^t \nabla\chi (B_s/R)\cdot a(B_s)ds
\rightarrow 0
\en
in probability, and then for some subsequence $R'''$ of $R''$, \kak{y14} converges to zero almost
surely. Thus $\int_0^t \chi_{R'''}(B_s) a(B_s) \circ dB_s \rightarrow \int_0^t  a(B_s) \circ dB_s$
almost surely,
and
\kak{y8} holds for any $a$ satisfying Assumption (A2).
\qed

\subsection{Functional integral representation}
Now we turn to constructing a functional integral representation for generalized
Schr\"o\-din\-ger operators including a vector potential term defined by \kak{bo3}.

A key element in our construction of a Feynman-Kac-type formula for $e^{-tH^\bern}$ is to
make use of a L\'evy subordinator.

\bt{3t}
Let $\bern\in\BERN$ and $V\in L^\infty(\BR)$. Under Assumption (A2) we have
\eq{bo8}
(f, e^{-tH^\bern   }g)= \ix \mathbb E^{x,0}_{P\times\nu} \lkkk \ov{f(B_0)}g(B_{T_t^\bern})
e^{-i\int_0^{T_t^\bern}\! a(B_s)\circ dB_s} e^{-\int_0^t
V(B_{T_s^\bern})ds} \rkkk.
\en
\et
\proof
We divide the proof into four steps. To simplify the notation, in this proof we drop the
superscript $\bern$ of the subordinator.

\vspace{0.1cm}
\noindent
\emph{(Step 1)} Suppose $V=0$. Then we claim that
\eq{bo88}
(f, e^{-t\bern(h)}g)= \ix \mathbb E^{x,0}_{P\times \nu} \lkkk \ov{f(B_0)}g(B_{T_t})
e^{-i\int_0^{T_t}\! a(B_s)\circ dB_s} \rkkk.
\en
To prove (\ref{bo88}) let $E^h$ denote the spectral projection of the self-adjoint operator
$h$. Then
\eq{bo12}
(f, e^{-t\bern(h) }g) = \int_{\Spec (h)} e^{-t\bern(u )} d(f, E_u ^h g).
\en
By inserting identity
(\ref{10}) in \kak{bo12} we obtain
$$
(f, e^{-t\bern(h) }g) = \int_{\Spec (h)} \EE_\nu^0[e^{-T_t u }]
d(f, E_u ^h g)=\EE_\nu^0 \lkkk (f, e^{-T_t h}g)\rkkk.
$$
Then by the Feynman-Kac-It\^o formula for $e^{-th}$ we have
$$
(f, e^{-t\bern(h) }g)=\EE_{\nu}^0\lkkk \ix \mathbb E^x_P\lkkk \ov{f(B_0)}g(B_{T_t})
e^{-i\int_0^{T_t}\! a(B_s)\circ dB_s}\rkkk\rkkk,
$$
thus \kak{bo88} follows.

\vspace{0.1cm}
\noindent
\emph{(Step 2)} Let  $0=t_0<t_1<\cdots <t_n$, $f_0,f_n\in \LR$ and assume that
$f_j\in L^\infty(\BR)$ for $j=1,...,n-1$. We claim that
\eq{bo10}
\lk f_0, \prod_{j=1}^n e^{-(t_j-t_{j-1})\bern(h) } f_j \rk
=\ix \mathbb E^{x,0}_{P\times\nu} \lkkk \ov{f_0(B_0)}\lk\prod_{j=1}^n f_j(B_{T_{t_j}})\rk
e^{-i\int_0^{T_t}\! a(B_s)\circ dB_s}\rkkk.
\en
For easing the notation write
$
G_j(\cdot)
=f_j(\cdot)
 \lk
 \prod_{i=j+1}^n e^{-(t_i-t_{i-1})\bern(h) }f_i\rk (\cdot)$.
By (Step 1) the left hand side of \kak{bo10} can be represented as
$$
\ix \mathbb E^{x,0}_{P\times\nu}\lkkk \ov{f_0(B_0)}
e^{-i\int_0^{T_{t_1-t_0}}\! a(B_s)\circ dB_s} G_1(B_{T_{t_1-t_0}})\rkkk.
$$
Let $\ms F^{\rm P}_t=\s(B_s,0\leq s\leq t)$ and $\ms F^\nu_t=\s(T_s,0\leq s\leq t)$ be the
natural filtrations. An application of the Markov property of $B_t$ yields
\begin{eqnarray*}
\lefteqn{
\lk f_0, \prod_{j=1}^n e^{-(t_j-t_{j-1})\bern(h) }f_j \rk}\\
&=&
\ix \mathbb E^{x,0}_{P\times\nu} \lkkk \ov{f_0(B_0)} e^{-i\int_0^{T_{t_1}}\!
a(B_s)\circ dB_s}
\EE_\nu^0 \EE_P^{B_{T_{t_1}}} \!\!\lkkk  f_1(B_0)e^{-i\int_0^{T_{t_2-t_1}}\! a(B_s)\circ dB_s}
G_2(B_{T_{t_2-t_1}}) \rkkk \rkkk\\
&=&
\ix \mathbb E^{x,0}_{P\times\nu} \lkkk \ov{f_0(B_0)} e^{-i\int_0^{T_{t_1}}\! a(B_s)\circ dB_s}
\right.\\
&&
\hspace{2cm} \left.
\EE_\nu^0 \lkkk \EE_P^0\lkkk \left. f_1(B_{T_{t_1}}) e^{-i\int_{T_{t_1}}^{T_{t_2-t_1}+T_{t_1}}\!
a(B_s)\circ dB_s}G_2(B_{T_{t_1}+T_{t_2-t_1}}) \right| {\ms F}^P_{T_{t_1}}\rkkk\rkkk\rkkk.
\end{eqnarray*}
Hence we obtain
\begin{eqnarray*}
&&
\hspace{-0.8cm}
\lk f_0, \prod_{j=1}^n e^{-(t_j-t_{j-1})\bern(h) }f_j \rk\\
&&
\hspace{-0.8cm}
= \ix \mathbb E^{x,0}_{P\times\nu} \!
\lkkk \ov{f_0(B_0)}e^{-i\int_0^{T_{t_1}}\!a(B_s)\circ dB_s}
 \EE_\nu^0\!\!
  \lkkk  f_1(B_{T_{t_1}})e^{-i\int_{T_{t_1}}^{T_{t_2-t_1}+T_{t_1}}\! a(B_s)\circ dB_s}
G_2(B_{T_{t_1}+T_{t_2-t_1}})
\rkkk\rkkk.
\end{eqnarray*}
The right hand side above can be rewritten as
$$
\ix \mathbb E^{x,0}_{P\times\nu}
\lkkk \ov{f_0(B_0)}e^{-i\int_0^{T_{t_1}}\! a(B_s)\circ dB_s}f_1(B_{T_{t_1}}) 
\EE^{T_{t_1}}_\nu\lkkk  e^{-i\int_{0}^{T_{t_2-t_1}}\! a(B_s)\circ dB_s}G_2(B_{T_{t_2-t_1}})
\rkkk\rkkk.
$$
Using now the Markov property of $T_t$ we see that
\begin{eqnarray*}
\lefteqn{
\lk f_0, \prod_{j=1}^n e^{-(t_j-t_{j-1})\bern(h) }f_j \rk} \\ &&
=\ix \mathbb E^{x,0}_{P\times\nu}\lkkk \ov{f_0(B_0)}e^{-i\int_0^{T_{t_1}}\! a(B_s)\circ dB_s}
f_1(B_{T_{t_1}})\mathbb E_\nu^0\lkkk \left. e^{-i\int_{T_{t_1}}^{T_{t_2}}\! a(B_s)\circ dB_s}
G_2(B_{T_{t_2}})\right|\ms F^\nu_{t_1}\rkkk\rkkk\\
&&
=\ix \mathbb E^{x,0}_{P\times\nu} \lkkk \ov{f_0(B_0)}e^{-i\int_0^{T_{t_1}}\! a(B_s)\circ dB_s}
f_1(B_{T_{t_1}})e^{-i\int_{T_{t_1}}^{T_{t_2}}\! a(B_s)\circ dB_s}G_2(B_{T_{t_2}})\rkkk.
\end{eqnarray*}
By the above procedure we obtain \kak{bo10}.

\vspace{0.1cm}
\noindent
\emph{(Step 3)} Suppose now that $0 \not= V\in L^\infty$ and it is continuous; we prove \kak{bo8}
for such $V$. Since $H^\bern$ is self-adjoint on $D(\bern (h))\cap D(V)$ the Trotter product
formula holds:
\begin{eqnarray*}
(f, e^{-t H^\bern}g) = \limn (f, (e^{-(t/n)\bern(h) }e^{-(t/n) V})^n g).
\end{eqnarray*}
(Step 2) yields
\begin{eqnarray*}
(f, e^{-t H^\bern}g)
&=&
\limn \ix \mathbb E^{x,0}_{P\times\nu} \lkkk \ov{f(B_0)}g(B_{T_t})
e^{-i\int_0^{T_t}\! a(B_s)\circ dB_s} e^{-\sum_{j=1}^n (t/n) V(B_{T_{tj/n}})}
\rkkk\\
&=&
\mbox{r.h.s. } \kak{bo10}
\end{eqnarray*}
Here we used that since $s\mapsto B_{T_s(\tau)}(\omega)$ has \emph{c\`adl\`ag} paths,
$V(B_{T_s(\tau)}(\omega))$ is continuous in $s\in [0,t]$ for each $(\omega,\tau)$ except for at most
finite points. Therefore $\sum_{j=1}^n \frac{t}{n} V(B_{T_{tj/n}}) \rightarrow \int_0^t V(B_{T_s})ds$
as $n\rightarrow \infty$ for each path and exists as a Riemann integral.

\vspace{0.1cm}
\noindent
\emph{(Step 4)}
An application of the method in \cite[Theorem 6.2]{sim04} will complete the proof of Theorem \ref{3t}.
To do that, suppose that $V\in L^\infty$ and $V_n=\phi(x/n)(V\ast j_n)$, where $j_n=n^d\phi(xn)$
with $\phi\in C_0^\infty(\BR)$ such that $0\leq \phi\leq 1$, $\int\phi(x)dx=1$ and $\phi(0)=1$.
Then $V_n(x)\rightarrow V(x)$ almost everywhere. $V_n$ is bounded and continuous, moreover $V_n(x)
\rightarrow V(x)$ as $n\rightarrow \infty$ for $x\not\in \ms N$, where the Lebesgue measure of $\ms
N$ is zero. Thus for almost every $(\omega,\tau)\in \BM\times \PO$, the measure of
$\{t\in [0,\infty)\,|\,B_{T_t(\tau)}(\omega)\in \ms N\}$ is zero.
Hence $\int_0^t V_n(B_{T_s})ds
\rightarrow \int_0^t V(B_{T_s})ds$ as $n\rightarrow \infty$ almost surely under $P^x\times \nu^0$,
\begin{eqnarray*}
\lefteqn{
\ix \mathbb E^{x,0}_{P\times\nu} \lkkk \ov{f(B_0)}g(B_{T_t})
e^{-i\int_0^{T_t}\! a(B_s)\circ dB_s} e^{-\int_0^t V_n(B_{T_s})ds} \rkkk}
\\ &&
\rightarrow \ix \mathbb E^{x,0}_{P\times\nu} \lkkk \ov{f(B_0)}g(B_{T_t})e^{-i\int_0^{T_t}\!
a(B_s)\circ dB_s} e^{-\int_0^t V(B_{T_s})ds} \rkkk
\end{eqnarray*}
as $n\rightarrow\infty$. On the other hand, $e^{-t(\bern(h)+V_n)}\rightarrow e^{-t(\bern(h)+ V)}$
strongly as $n\rightarrow \infty$, since $\bern(h)+V_n$ converges to $\bern(h)+V$ on the common
domain $D(\bern(h))$. Thus the theorem follows.
\qed

Setting $a=0$ and $\bern(u)=u^\alpha$ yields an interesting class of its own.
\begin{definition}[\textbf{Fractional Schr\"odinger operator}]
{\rm
Let $0<\alpha < 1$ and $\bern(u) = u^\alpha$. We call
\begin{equation}
H_{\alpha} =\left(\frac{1}{2}{\rm p}^2 \right)^{\alpha} + V
\end{equation}
\emph{fractional Schr\"odinger operator} with exponent $\alpha$.
}\end{definition}

\begin{corollary}[\textbf{Functional integral for fractional Schr\"odinger operator}]\ \\
Let $T_t^\bern$ be the subordinator for a fractional Schr\"odinger operator, i.e., an
$\alpha$-stable process, and $V\in L^\infty(\BR)$. Then
$$
(f, e^{-tH_\alpha }g)= \ix \mathbb E^{x,0}_{P\times\nu} \lkkk \ov{f(B_0)}g(B_{T_t})
e^{-\int_0^t V(B_{T_s^\bern})ds} \rkkk.
$$
\end{corollary}

We use the notation $\is T=\inf\Spec{T}$ here and in Sections 5 and 6 below.
\begin{corollary}
[\textbf{Diamagnetic inequality}]
\label{wr2}
Let $\bern\in \BERN $, $V\in L^\infty(\BR)$, and Assumption (A2) hold. Then
\eq{dia1}
|(f,e^{-tH^\bern  }g)|\leq (|f|, e^{-t(\bernz +V)}|g|)
\en
and the energy comparison inequality
\eq{dia2}
\is{\bernz +V} \leq \is {H^\bern}
\en
holds.
\label{dia}
\end{corollary}
\proof
By Theorem \ref{3t} we have
$$
|(f,e^{-tH^\bern  }g)|\leq \ix  \mathbb E_{P\times\nu}^{x,0}
\lkkk
|f(B_0)| |g(B_{T_t^\bern})| e^{-\int_0^t V(B_{T_s^\bern}) ds}
\rkkk.
$$
The right hand side above coincides with that of \kak{dia1}, and \kak{dia2} follows directly
from \kak{dia1}.
\qed

\subsection{Singular external potentials}
By making use of the functional integral representation obtained in the previous subsection we
can now also consider more singular external potentials.
\bt{e}
Let Assumption (A2) hold.
\begin{itemize}
\item[(1)]
Suppose $|V|$ is relatively form bounded with respect to $\bernz $ with relative bound $b$.
Then $|V|$ is also relatively form bounded with respect to $\bern(h)$ with a relative bound
not larger than $b$.
\item[(2)]
Suppose $|V|$ is relatively bounded with respect to $\bernz $ with relative bound $b$. Then
$|V|$ is also relatively bounded with respect to $\bern(h)$ with a relative bound not larger
than $b$.
\end{itemize}
\et
\proof
The proof is parallel with
that of \cite[Theorem 15.6]{sim04}.
By virtue of Corollary \ref{wr2}
we have
\eq{bsi}
|(f, e^{-t\bern(h)}g)|\leq (|f|, e^{-t\bernz}|g|).
\en
Since $\d (\bern(h)+E)^{-\han}=\frac{1}{\sqrt\pi}\int_0^\infty {t^{-\han}} e^{-t(\bern(h)+E)}dt$,
$E>0$, \kak{bsi} implies that
\eq{ko19-1}
\left|(\bern(h)+E)^{-\han}f\right|(x) \leq (\bernz +E)^{-\han}|f|(x)
\en
for almost every $x\in\BR$. Hence we have
$$
|V(x)|^\han \left|(\bern(h)+E)^{-\han}f\right|(x) \leq |V(x)|^\han \bernz +E)^{-\han}|f|(x)
$$
and
\eq{ko19}
\frac{\||V|^{\han}(\bern(h)+E)^{-\han}f\|}{\|f\|}
\leq
\frac{\||V|^\han (\bernz +E)^{-\han}|f|\|}{\|f\|}.
\en
Similarly, by using $\d (\bern(h)+E)^{-1} = \int_0^\infty e^{-t(\bern(h)+E)}dt$, $E>0$,
we have
\eq{ko20}
\frac{\||V|(\bern(h)+E)^{-1}f\|}{\|f\|}
\leq
\frac{\||V| (\bernz +E)^{-1}|f|\|}{\|f\|}.
\en
On taking the limit $E\rightarrow\infty$, the right hand sides of \kak{ko19} and \kak{ko20}
converge to $b$; compare \cite[Lemma 13.6]{hs}, \cite{sim04, ahs}. Hence (1) follows by
\kak{ko19} and (2) by \kak{ko20}.
\qed
\bc{ko21}
(1) Take Assumption (A2) and let $V$ be relatively bounded with respect to $\bernz$ with
relative bound strictly smaller than one. Then $\bern (h)+V$ is self-adjoint on $D(\bern (h))$
and bounded from below. Moreover, it is essentially self-adjoint on any core of $\bern (h)$.
(2) Suppose furthermore (A3). Then $C_0^\infty(\BR)$ is an operator core of $\bern (h)+V$.
\ec
\proof
(1) By (2) of Theorem \ref{e}, $V$ is relatively bounded with respect to $\bern (h)$ with a
relative bound strictly smaller than one. Then the corollary follows by the Kato-Rellich
theorem. (2) follows from Theorem \ref{main2}.
\qed

Theorem \ref{e} also allows $\bern(h)+V$ to be defined in form sense. Let $V=V_+-V_-$ where
$V_+=\max\{V,0\}$ and $V_-=\min\{-V,0\}$. Theorem \ref{e} implies that whenever $V_-$ is form bounded
to $\bernz $ with a relative bound strictly smaller than one, it is also form bounded with respect to
$\bern(h)$ with a relative bound strictly smaller than one. Moreover, assume that $V_+\in L_{\rm loc}^1(\BR)$.
We see that given Assumption (A1), $Q(\bern(h))\cap Q(V_+) \supset C_0^\infty(\BR)$ by Corollary \ref{ko21}.
In particular, $Q(\bern(h))\cap Q(V_+)$ is dense. Define the quadratic form
\eq{ko22}
{\rm q} (f,f):= (\bern(h)^\han f, \bern(h)^\han f)+(V_+^\han f,V_+^\han f)-(V_-^\han f,V_-^\han f)
\en
on $Q(\bern(h) )\cap Q(V_+)$. By the KLMN Theorem \cite{rs2} ${\rm q}$ is a semibounded closed form.

\begin{definition}
{\bf (Generalized Schr\"odinger operator with singular $V$)}
\label{ko1}
{\rm
Let Assumption (A2) hold and $V=V_+-V_-$ be such that $V_+\in \loc 1$ and $V_-$ is form bounded with respect
to $\bern(\half{\rm p}^2)$ with a relative bound strictly less than 1. We denote the self-adjoint operator
associated with \kak{ko22} by $\bern(h) \, \, \dot+\,\,  V_+\, \, \dot-\, \,  V_-$ defined as a quadratic
form sum.
}
\end{definition}
Since we need (A2) to show the relative form boundedness of $V_-$ with respect to $\bern (h)$,
 (A2) is assumed in Definition \kak{ko1}.
 \noindent

Now we are in the position to extend Theorem \ref{3t} to potentials expressed as form sums.
\bt{f}
Take Assumption (A2). Let $V=V_+-V_-$ be such that $V_+\in \loc 1$ and $V_-$ is infinitesimally
small with respect to $\bern(\half{\rm p}^2)$ in form sense. Then the functional integral
representation given by Theorem \ref{3t} also holds for $\bern(h) \, \, \dot+\, \,  V_+\, \,
\dot- \, \,  V_-$.
\et
\proof
Write
\begin{eqnarray*}
V_{+,n}(x) =
\left\{
\begin{array}{ll}
V_+(x),&V_+(x)<n,\\
n,& V_+(x) \geq n,
\end{array}
\right.\quad
V_{-,m}(x)=
\left\{
\begin{array}{ll}
V_-(x),&V_-(x)<m,\\
m,& V_-(x) \geq m.
\end{array}
\right.
\end{eqnarray*}
For simplicity we write just $\bern$ for $\bern(h)$. Define the closed quadratic forms
\begin{eqnarray*}
{\rm q}_{n,m}(f,f)
&=&
(\bern^\han f, \bern^\han f) + (V_{+,n}^{\han}f,V_{+,n}^{\han}f)-(V_{-,m}^{\han}f,V_{-,m}^{\han}f), \\
{\rm q}_{n,\infty}(f,f)
&=&
(\bern^\han f,\bern^\han f)+(V_{+,n}^{\han}f,V_{+,n}^{\han}f)-(V_{-}^{\han}f,V_{-}^{\han}f), \\
{\rm q}_{\infty, \infty}(f,f)
&=&
(\bern^\han f,\bern^\han f)+(V_{+}^{\han}f,V_{+}^{\han}f)-(V_{-}^{\han}f,V_{-}^{\han}f),
\end{eqnarray*}
where the form domains are given by
$$
Q({\rm q}_{n,m})=Q(\bern ),\quad Q({\rm q}_{n,\infty})=Q({\bern }), \quad
Q({\rm q}_{\infty, \infty})=Q({\bern })\cap Q(V_+).
$$
Clearly, ${\rm q}_{n,m}\geq {\rm q}_{n,m+1}\geq {\rm q}_{n,m+2}\geq...\geq {\rm q}_{n,\infty}$ and
${\rm q}_{n,m}\rightarrow {\rm q}_{n,\infty}$ in the sense of quadratic forms on $\cup_m Q({\rm q}_{n,m})=
Q({\bern })$. Since ${\rm q}_{n,\infty}$ is closed on $Q({\bern })$, by the monotone convergence theorem
for a non-increasing sequence of forms (see \cite[Theorem VIII.3.11]{kat76} and \cite[Theorem 3.2]{sim78})
the associated positive self-adjoint operators satisfy $\bern\, \, \dot{+}\, \, V_{+,n}\, \, \dot{-}\, \,
V_{-,m} \rightarrow \bern \, \, \dot{+}\, \, V_{+,n}\, \, \dot{-}\, \, V_{-}$ in strong resolvent sense,
which implies that
\eq{mono1}
e^{-t\left({\bern }
\, \, \dot{+}\, \, V_{+,n}\, \, \dot{-}\, \, V_{-,m}\right)}
\rightarrow
e^{-t \left({\bern }
\, \, \dot{+}\, \, V_{+,n}\, \, \dot{-}\, \, V_{-}\right)}
\en
strongly as
$m\rightarrow \infty$,
for all $t\geq0$. Similarly, we have ${\rm q}_{n,\infty}\leq {\rm q}_{n+1,\infty}
\leq {\rm q}_{n+2,\infty}\leq ...\leq {\rm q}_{\infty,\infty}$ and ${\rm q}_{n,\infty}\rightarrow
{\rm q}_{\infty,\infty}$ in quadratic form sense on $\{f\in\cap_n Q({\rm q}_{n,\infty}) \,|\,\sup_n
{\rm q}_{n,\infty}(f,f)<\infty \} = Q({\bern })\cap Q(V_+)$. Hence by the monotone convergence
theorem for a non-decreasing sequence of forms (see \cite[Theorem 3.1 and Theorem 4,1]{sim78} and
\cite[Theorem VIII.3.13 with Supplementary notes to Chapter VIII,5 (p.575)]{kat76}) we obtain
\eq{mono2}
e^{-t \left({\bern }\,\, \dot{+}\,\, V_{+,n}\,\,\dot{-}\,\, V_-\right)} \rightarrow
e^{-t \left({\bern }\, \, \dot{+}\, \, V_{+}\, \, \dot{-}\, \, V_{-}\right)},
\en
for all $t\geq0$, in
strong sense as $n\rightarrow\infty$.

On the other hand, we look at the convergence of the expression
\eq{kieta}
\int \!dx\EE_{P\times\nu}^{x,0}
\lkkk   e^{-\int_0^t(V_{+,n}-V_{-,m})(B_{T_s^\bern}) ds} I\rkkk.
\en
Here $I=\ov{f(B_0)}e^{-i\int_0^{T_t}a(B_s)\circ dB_s}g(B_{T_t})$. Decompose $I$ into its real and
imaginary parts, and further into their positive and negative parts $\Re I=\Re I_+-\Re I_-$ and
$\Im I=\Im I_{+}-\Im I_{-}$. Then by \kak{mono1} and the monotone convergence theorem
\begin{eqnarray*}
\int \!dx\EE_{P\times\nu}^{x,0} \lkkk  e^{-\int_0^t(V_{+,n}-V_{-,m})(B_{T_s^\bern}) ds} \Re I_+ \rkkk
\longrightarrow
\int \!dx\EE_{P\times\nu}^{x,0} \lkkk   e^{-\int_0^t(V_{+,n}-V_{-})(B_{T_s^\bern}) ds} \Re I_+ \rkkk
\end{eqnarray*}
as $m\rightarrow\infty$. Similarly, the remaining three terms $\Re I_-$, $\Im I_+$ and $\Im I_-$
also converge. Thus \kak{kieta} converges to $\d \int \!dx\EE_{P\times\nu}^{x,0} \lkkk
e^{-\int_0^t(V_{+,n}-V_{-})(B_{T_s^\bern}) ds} I \rkkk$ as $m\rightarrow \infty$. Moreover,
$$
\int \!dx\EE_{P\times\nu}^{x,0} \lkkk   e^{-\int_0^t(V_{+,n}-V_{-})(B_{T_s^\bern}) ds} I \rkkk
\longrightarrow \int \!dx\EE_{P\times\nu}^{x,0} \lkkk   e^{-\int_0^t(V_{+}-V_{-})(B_{T_s^\bern}) ds}
I\rkkk
$$
as $n\rightarrow\infty$, by \kak{mono2} and the dominated convergence theorem. Thus the proof is
complete.
\qed

\section{$\bern$-Kato class potentials}
\subsection{Definition of $\bern$-Kato class potentials}
In this section we give a meaning to Kato class  for potentials $V$ relative to $\bern$ and
extend generalized Schr\"odinger operators with vector potential to such $V$.

It is known that the composition of a Brownian motion and a subordinator yields a L\'evy process.
Recall that for given $\bern \in \BERN$, the random process
\eq{random}
X_t:\BM\times \SU \ni (\omega,\tau) \mapsto B_{T^\bern _t(\tau)}(\omega)
\en
is called $d$-dimensional subordinated Brownian motion with respect to the subordinator $(T^\bern _t)_
{t\geq 0}$. It is a L\'evy process whose properties are determined by the pair $(b,\la )$ in (\ref{6}).
Its characteristic function is
\begin{equation}
\EE_{P\times \nu}^{0,0}[e^{i\xi\cdot X_t}]=e^{-t\bern(\xi\cdot\xi/2)},\quad \xi\in\BR.
\label{cha}
\end{equation}
\begin{assumption}
\label{a}
Let $\bern \in \BERN$ be such that
\eq{bound1}
\int_\BR e^{-t\bern(\xi\cdot\xi/2)}d\xi<\infty
\en
for all $t > 0$.
\end{assumption}

Let $\bern\in\BERN$ and $(b,\la)\in \RR_+\times \ms L$ be its corresponding non-negative drift coefficient and
L\'evy measure, i.e., $\bern(u)=bu+\int _0^\infty \lk 1-e^{-uy}\rk \la(dy)$. It is clear that if $b>0$, then
\kak{bound1} is satisfied. In the case of $b=0$ but $\int_0^1\la(dy)<\infty$, since $\sup_{u\geq 0} \bern(u)<
\infty$, \kak{bound1} is not satisfied. Thus $\bern$ obeying \kak{bound1} at least satisfies $\int_0^1\la(dy)=
\infty$ when $b=0$. In this case we have
$$
\bern (u^2/2)\geq \int_0^1 (1-e^{-u^2 y/2})\la(dy) \geq (1-e^{-1}) \int_0^1 (\frac{u^2 y}{2}\wedge  1) \la(dy)
\geq
(1-e^{-1}) \int_{2/u^2}^1 \la(dy).
$$
Thus in case $b=0$ and $\int_0^1\la(dy)=\infty$, assuming that there exists $\rho(u)$ such that $\int_{2/u^2}^1
\la(dy)\geq \rho(u)$ and $\int_\BR e^{-t \rho(|\xi|)}d\xi<\infty$, we can make sure Assumption \ref{a} holds.

Under Assumption \ref{a} we define
\begin{equation}
p_t(x)=\frac{1}{(2\pi)^d}
\int_\BR e^{-ix\cdot \xi}e^{-t\bern(\xi\cdot\xi /2)}d\xi
\label{pt}
\end{equation}
and
$$
\Pi_\la (x)=\int_0^\infty e^{-\la t} p_t(x) dt.
$$
$p_t(x)$ denotes the distribution density of $X_t$ in \kak{random} and $\Pi_\la(x-y)$ is the integral kernel of
the resolvent $(\bernz +\la)^{-1}$ with $\la>0$, i.e.,
$$
\lk f, \lk \bernz +\la\rk^{-1}g\rk =\int_{\BR\times\BR} \overline{f(x)} g(y) \Pi_\la(x-y) dxdy.
$$
Clearly, $p_t(x)$ and $\Pi_\la(x)$ are spherically symmetric. For $f\in C_0^\infty(\BR)$ it follows that \eq{tr2}
\EE_{P\times\nu}^{0,0}[f(X_t)]=\int f(x) p_t(x) dx.
\en
Hence for non-negative $f\in C_0^\infty(\BR)$, the right hand side of \kak{tr2} is non-negative since so is
the left hand side. Thus $p_t(x)\geq 0$ for almost every $x\in\BR$. By a limiting argument with $f\rightarrow 1$,
we can also see that $p_t\in L^1(\BR)$ and $\|p_t\|_{L^1(\BR)}=1$ by \kak{tr2}.

We moreover compute $\Pi_\la$ as
\begin{eqnarray*}
\Pi_\la(x) = (2\pi)^{-d/2}\frac{1}{|x|^{(d-1)/2}}\int_0^\infty \frac{r^{(d-1)/2}}{\la+\bern(r^2/2)}
\sqrt{r|x|} J_{(d-2)/2}(r|x|) dr,
\end{eqnarray*}
with the Bessel function given by
$$
\d J_\nu(s)=
\lk\frac{s}{2}\rk^\nu
\frac{1}{
\sqrt\pi \Gamma(\nu+\half)}\int_0^\pi e^{is \cos \theta}(\sin\theta)^{2\nu}d\theta
=\sum_{n=0}^\infty
\frac{(-1)^n}{n!\Gamma(n+\nu+1)}
\lk\frac{s}{2}\rk^{2n+\nu}.
$$
Note that
$\sup_{u\geq0}\sqrt{u}J_\nu(u)<\infty$.

Let
$$\|f\|_{l^1(L^\infty)}=\sum_{\alpha\in \mathbb Z^d}\sup_{x\in C_\alpha}|f(x)|,$$
where $C_\alpha$
denotes the unit cube centered at $\alpha\in \mathbb Z^d$.
We introduce an additional assumption on distribution density $p_t$.
\begin{assumption}\label{b}
Let $p_t$ be such that $\sup_{t>0}\|1_{\{|x|>\delta\}}p_t\|_{l^1(L^\infty)}<\infty.$
\end{assumption}
Let $f$ be a real valued function on $\BR$. When $r\mapsto f(rx)$ is non-increasing on $[0,\infty)$,
we say that $f$ is radially non-increasing. In $d=1$ for a radially non-increasing $L^1$-function $f$
it can be seen by the definition of $l^1(L^\infty)$ that there exists a constant $C_\delta=C_\delta(f)$
such that
\eq{tr}
\|1_{\{|x|>\delta\}}f\|_{l^1(L^\infty)}\leq C_\delta \|f\|_{L^1}.
\en
In the general case $d\geq 2$ it can be also seen that \kak{tr} holds for all radially non-increasing $f$,
see \cite[p.~131, Corollary]{CMS90}. In particular, Assumption \ref{b} is satisfied whenever $p_t$ is
radially non-increasing, since $\|p_t\|_{L^1}=1$.

\begin{example}
{\bf ($\alpha$-stable subordinator)}
{\rm
In the case of $\bern(u)=u^\alpha$, $0\leq \alpha\leq 1$, it is clear that Assumption \ref{a} is satisfied.
It is also known that the distribution density $p_t^\alpha $ of $B_{T_t^\bern}$ is radially non-increasing.
This is proven by a unimodality argument of spherically symmetric distribution functions; see
\cite[Theorem 4.1]{kan77}, \cite[Theorem 2]{wol78}, \cite[p.132]{CMS90}, \cite[Theorem 1]{yam78}, and
\cite{Sato99} for details on unimodality. Then Assumption \ref{b} is again satisfied.
}
\end{example}

\begin{example}
{\rm
Let $\bern(u)=\sqrt{2u+m^2}-m$, $m\geq0$. It is clear that Assumption \ref{a} is satisfied. The distribution
function $p_t$ of $B_{T_t^\bern}$ is expressed as
$$
p_t(x)=(2\pi)^{-d} \frac{1}{\sqrt{|x|^2+t^2}} \int_{\BR} e^{mt} e^{-\sqrt{(|x|^2+t^2)(p^2+m^2)}}dp,
$$
see \cite[(2.7)]{hsl}. Then $p_t$ is indeed radially non-increasing.
}
\end{example}

The next proposition allows an extension of $\bernz $ to Kato class.
\begin{proposition}
\label{d}
Let $V\geq 0$. Under Assumptions \ref{a} and \ref{b} the following three properties are equivalent:
\begin{itemize}
\item[(1)]
$\d \lim_{t\downarrow 0}\sup_{x\in\RR^d}\int_0^t \EE_{P\times\nu}^{x,0}[V(X_s)]ds=0$,
\item[(2)]
$\d \lim_{\la \rightarrow \infty}\sup_{x\in \RR^d}\lk(\bernz +\la )^{-1}
V\rk(x)=0$,
\item[(3)]
$\d \lim_{\delta \downarrow 0}\sup_{x\in\RR^d}\int_{|x-y|<\delta}\Pi_1(x-y)V(y)dy=0$.
\end{itemize}
\ep
\proof
Similar to Theorem III.1 in \cite{CMS90}.
\qed

\begin{definition}[\textbf{$\bern$-Kato class}]
\rm{
Take Assumptions \ref{a} and \ref{b}. Write $V=V_+-V_-$ in terms of its positive and negative parts.
The \emph{$\bern$-Kato class} is defined as the set of potentials $V$ for which $V_-$ and $1_C V_+$
with every compact subset $C \subset {\mathbb  R}^d$ satisfy any of the three equivalent conditions in
Proposition \ref{d}. Here $1_C$ denotes the indicator function on $C$.
\label{katoclass}
}
\end{definition}

By (3) of Proposition \ref{d} we can derive explicit conditions defining $\bern$-Kato class using
the relation of the L\'evy measure of the subordinator with the associated Bernstein function.
\begin{example}
In the case $d=3$, since $J_{1/2}(x)=(2/\pi)^\han x^{-1/2} \sin x$, we have
$$
\Pi_\la(x)=\frac{1}{2\pi^2|x|} \int_0^\infty \frac{r\sin r}{|x|^2 \lk
\la+\bern\lk \frac{r^2}{2|x|^2}\rk \rk }dr.
$$
\end{example}

\begin{example}
\rm{
Let $T_t$ be an $\alpha$-stable process generated by $((\han){\rm p}^2)^\alpha$, $\alpha
\in (0,1)$. Then a calculation gives that $V$ is in $\bern$-Kato class in the sense of
Definition \ref{katoclass} if and only if
\begin{eqnarray*}
\Pi_1(x) =
\left\{
\begin{array}{ll}
\d c(d,2\alpha)|x|^{2\alpha-d}, &  \;  2\alpha < d;\\
\d -\frac{1}{\pi} \log  |x|,                     &  \;  2\alpha = d \in \{1,2\};\\
\d c(1,2\alpha)|x|^{2\alpha-1}, &  \;  2\alpha > d = 1,
\end{array}
\right.
\end{eqnarray*}
where
$
\d c(d,\beta) := \frac{\Gamma\left((d-\beta)/2\right)}{2^\beta \pi^{d/2}|\Gamma(\beta/2)|}.
$
}
\end{example}
\begin{remark}
\rm{
For $\bern$-Kato class potentials $V$ condition (2) of Proposition \ref{d} implies that $V_-$ is
infinitesimally form bounded with respect to $\bernz $. In this case $\bernz +V$ can be defined
in form sense.
}
\end{remark}

\subsection{$\bern$-Kato class potential and hyper\-contractivity}
In this section we construct Schr\"odinger semigroups with $\bern$-Kato class potentials
and show their hypercontractivity property. References on the hypercontractivity for semigroups
with usual Schr\"odinger operators with magnetic field include \cite{sim82, BHL00}.

\bl{muyami}
Let $V\geq 0$ and $\bern\in \BERN$. Suppose that $V$ satisfies (1) of Proposition \ref{d}. Then for
$t\geq 0$,
\eq{ka1}
\sup_{x\in\BR}\EE_{P\times\nu}^{x,0}\left[e^{\int_0^t V(X_s) ds}\right]<\infty.
\en
\el
\proof
There exists $s>0$ such that $\sup_{x\in\BR} \EE_{P\times \nu}^{x,0} [\int_0^s V(X_s)ds]=\epsilon<1$
by (1) of Proposition \ref{d}. Then by the Khas'minskii Lemma we conclude that
$$
\sup_{x\in\BR} \EE_{P\times \nu}^{x,0} \lkkk e^{\int_0^s V(X_s)ds}\rkkk\leq (1-\epsilon)^{-1}.
$$
Consider the image measure $\rho$ of $(X_t)_{t\geq 0}$ on the space $D([0,\infty);\BR)$ of {\it c\'adl\'ag}
paths. Then $\mathbb E_\rho^x \left[e^{\int_0^sV(X_s) ds}\right]=\EE_{P\times \nu}^{x,0}
\left[e^{\int_0^sV(X_s) ds}\right]$ and clearly $(X_t)_{t\geq 0}$ is a Markov process with respect to
$\rho$. Furthermore,
\begin{eqnarray*}
\mathbb E_\rho^x
\left[e^{\int_0^{2s} V(X_s)ds}\right]
&=&
\mathbb E_\rho^x
\left[e^{\int_0^{s} V(X_s)ds}e^{\int_s^{2s} V(X_s)ds}\right] \\
&=&
\mathbb E_\rho^x
\left[e^{\int_0^{s} V(X_s)ds} \EE_\rho^{X_s}[e^{\int_0^s V(X_s)ds}] \right] \\
&\leq&
\lk \sup_{y\in\BR} \mathbb E_\rho^y [e^{\int_0^{s} V(X_s)ds}]\rk \mathbb E_\rho^x
[e^{\int_0^{s} V(X_s)ds}]\\
&\leq&
(1-\epsilon)^{-2}.
\end{eqnarray*}
Repeating this procedure we obtain \kak{ka1} for all $t\geq 0$.
\qed

The next result says that we can define a Feynman-Kac semigroup for $\bern$-Kato class potentials.
\noindent
\bt{k5}
Let $\bern\in\BERN$, $V$ belong to $\bern$-Kato class and let Assumption (A2) hold. Consider
$$
{\rm U}_t f(x)=\mathbb E^{x,0}_{P\times\nu}\lkkk e^{-i\int_0^{T_t^\bern }a(B_s)\circ dB_s}
e^{-\int_0^t V(B_{T_s^\bern})ds}f(B_{T_t^\bern})\rkkk.
$$
Then ${\rm U}_t $ is a strongly continuous symmetric semigroup. In particular, there exists a
self-adjoint operator $K^\bern$ bounded from below such that ${\rm U}_t = e^{-tK^\bern}$.
\et
\proof
Let $V=V_+-V_-$. Hence by Lemma \ref{muyami} we have
\begin{eqnarray*}
\| {\rm U}_t f\|^2
&\leq &
\ix  \mathbb E _{P\times \nu}^{x,0}\lkkk e^{-2\int_0^t V_+(X_s)ds}|f(X_t)|^2\rkkk
\mathbb E _{P\times \nu}^{x,0}\lkkk e^{2\int_0^t V_-(X_s)ds}\rkkk\\
&\leq&
C_t \ix  \mathbb E _{P\times \nu}^{x,0}\left|f(X_t)|^2 \rkkk \\
&=&
C_t \|e^{-t\bernz }f\|^2 \leq
C_t\|f\|^2,
\end{eqnarray*}
where $C_t=\sup_{x\in\BR}\EE_{P\times \nu}^{x,0}[e^{2\int_0^tV_-(X_s)ds}]$. Thus ${\rm U}_t $ is a
bounded operator from $\LR$ to $\LR$. In the same manner as in Step~2 of the proof of Theorem \ref{3t}
we conclude that the semigroup property ${\rm U}_t {\rm U}_s={\rm U}_{t+s}$ holds for  $t,s\geq 0$. We
check strong continuity of ${\rm U}_t $ in $t$; it suffices to show weak continuity. Let $f,g\in
C_0^\infty(\BR)$ and simply we write $T_t$ for $T_t^\bern$. Then we have
$$
(f,{\rm U}_t g)=\ix  \mathbb E_{P\times\nu}^{x,0} \lkkk \ov{f(B_0)} g(B_{T_t})
e^{-i\int_0^{T_{t}} a(B_s)\circ dB_s} e^{-\int_0^t V(B_{T_s}) ds} \rkkk.
$$
Since $T_t(\tau)\rightarrow 0$ as $t\rightarrow 0$ for each $\tau\in \SU$, the dominated convergence
theorem gives $(f,{\rm U}_t g)\rightarrow (f,g)$.

Finally we check the symmetry property ${\rm U}_t^\ast ={\rm U}_t$. By a limiting argument it is
enough to show this for $a\in (C_{\rm b}^2(\BR))^d$. Let $\widetilde B_s=\widetilde B_s(\omega,\tau)=
B_{T_t(\tau)-s}(\omega)-B_{T_t(\tau)}(\omega)$. Then for each $\tau\in \SU $, $\widetilde B_s\stackrel
{\rm d}{=} B_s$ with respect to $dP^x$. (Here $ Z\stackrel{\rm d}{=} Y$ denotes that $Z$ and $Y$ are
identically distributed.) Thus there exists a sequnece $\{n\}\subset \mathbb N$ such that
\begin{eqnarray*}
(f, {\rm U}_t g)
&=&
\EE_{P\times\nu}^{0,0} \left[\ix \overline{f(x)} e^{-i\int_0^{T_t}a(x+\widetilde B_s)\circ d\widetilde B_s}
e^{-\int_0^t V(x+\widetilde B_{T_s})} g(x+\widetilde B_{T_t}) \right]\\
&=&
\lim_{n\to\infty}
\EE_{P\times\nu}^{0,0} \left[\ix \overline{f(x)} e^{-i\sum_{j=1}^n I_j}
e^{-\int_0^t V(x+\widetilde B_{T_s})} g(x+\widetilde B_{T_t}) \right],
\enn
where
$I_j=
\half
\lk
a(x+\widetilde
B_{T_tj/n})+
a(x+\widetilde B_{T_t(j-1)/n})\rk
(B_{T_tj/n}-
B_{T_t(j-1)/n})
$.
Changing the variable $x$ to $y=x+\widetilde B_{T_t}$,
 we have
\begin{eqnarray*}
(f, {\rm U}_t g)
=
\lim_{n\to\infty}
\EE_{P\times\nu}^{0,0} \left[\int_\BR\!\!\! dy \overline{f(
y-\widetilde B_{T_t})}
e^{-i\sum_{j=1}^n \widetilde I_j}
e^{-\int_0^t V(y-\widetilde B_{T_t}+\widetilde B_{T_s})} g(y) \right],
\enn
where
$$\widetilde I_j=
\half
\lk
a(y-\widetilde B_{T_t}+\widetilde
B_{T_tj/n})+
a(y-\widetilde B_{T_t}+\widetilde B_{T_t(j-1)/n})
\rk
(\widetilde B_{T_tj/n}-
\widetilde B_{T_t(j-1)/n})
.$$
Since $\widetilde B_{T_s}-\widetilde B_{T_t}\stackrel{\rm d}{=}B_{T_t-T_s}$,
  we can compute $
\lim_{n\to\infty}\sum_{j=1}^n \widetilde I_j$
 in $L^2(\BM,dP^0)$
as
\begin{eqnarray*}
&&
\lim_{n\to\infty}
\sum_{j=1}^n \widetilde I_j\\
&&=
\lim_{n\to\infty}
\sum_{j=1}^n
\half
\lk
a(y+B_{T_t-T_tj/n})
+a(y+B_{T_t-T_t(j-1)/n})
\rk
(B_{T_t-T_tj/n}-
B_{T_t-T_t(j-1)/n})\\
&&=-
\lim_{n\to\infty}
\sum_{j=1}^n
\half
\lk
a(y+B_{T_tj/n})
+a(y+B_{T_t(j-1)/n})
\rk
(B_{T_tj/n}-
B_{T_t(j-1)/n})\\
&&=-\int_0^{T_t}a(B_s)\circ dB_s.
\enn
Then
we have
\begin{eqnarray*}
(f, {\rm U}_t g)= \ix \EE_{P\times\nu}^{x,0} \left[\ov{f(B_{T_t})}
e^{+i\int_0^{T_t} a(B_s) \circ dB_s} e^{-\int_0^tV(B_{T_t-T_s})ds} g(x) \right].
\end{eqnarray*}
Moreover, as $T_t-T_s\stackrel{\rm d}{=}T_{t-s}$ for $0\leq s\leq t$, we obtain
\begin{eqnarray*}
(f, {\rm U}_t g)
&=&
\ix \EE_{P\times\nu}^{x,0} \left[\ov{f(B_{T_t})}
e^{+i\int_0^{T_t} a(B_s) \circ dB_s} e^{-\int_0^t V(B_{T_{t-s}})ds }g(x) \right]\\
&=&
\ix \ov{\EE_{P\times\nu}^{x,0} \left[f(B_{T_t})
e^{-i\int_0^{T_t} a(B_s) \circ dB_s} e^{-\int_0^t V(B_{T_s})ds }\right]} g(x)
\\
&=&
({\rm U}_t f, g).
\end{eqnarray*}
The existence of a self-adjoint operator $K^\bern$ bounded from below such that ${\rm U}_t
= e^{-tK^\bern}$ is a consequence of the Hille-Yoshida theorem. This completes the
proof.
\qed

\begin{definition}[\textbf{$\bern$-Kato class Schr\"odinger operator}]
{\rm
Let $V$ be in $\bern$-Kato class and take Assumption (A2). We call $K^\bern$ given in Theorem
\ref{k5} \emph{generalized Schr\"odinger operator for $\bern$-Kato class potentials}. We refer
to the one-parameter operator semigroup $e^{-tK^\bern}$, $t\geq 0$, as the \emph{$\bern$-Kato
class generalized Schr\"odinger semigroup}.
}
\end{definition}
Put $K^\bern_0$ for the operator defined by $K^\bern $ with $a$ replaced by $0$.

\bt{hyp}{\rm(\textbf{Hypercontractivity})}
Let $V$ be a $\bern$-Kato class potential and assume (A2) to hold. Then $e^{-tK^\bern}$ is a
bounded operator from $L^p(\BR)$ to $L^q(\BR)$, for all $1\leq p\leq q\leq \infty$. Moreover,
$\|e^{-tK^\bern }\|_{p,q}\leq \|e^{-tK^\bern_0}\|_{p,q}$ holds for all $t\geq 0$.
\et
\proof
By the Riesz-Thorin theorem it suffices to show that $e^{-tK^\bern}$ is bounded as
an operator of
(1) $L^\infty(\BR)\rightarrow L^\infty(\BR)$, (2) $L^1(\BR)\rightarrow L^1(\BR)$ and (3) $L^1(\BR)
\rightarrow L^\infty(\BR)$. Since
\eq{huhigi}
|e^{-tK^\bern}f(x)| \leq e^{-tK^\bern_0}|f|(x),
\en
we will prove (1)-(3) for $e^{-tK^\bern_0}$. For simplicity we denote $\EE_{P\times\nu}^{x,0}=\EE^x$
and ${\rm P}_t=e^{-tK^\bern_0}$, i.e., we have
$$
{\rm P}_tf(x)=\EE^x[e^{-\int_0^t V(X_s)ds}f(X_t)].
$$
To consider (1), let $f\in L^\infty(\BR)$. We have by Lemma \ref{muyami},
$$
\|{\rm P}_tf\|_\infty \leq \sup_{x\in\BR} \lk \EE^x[e^{-\int_0^t V(X_s) ds}]\rk \|f\|_\infty.
$$
Thus (1) follows.

To derive (2), let $0\leq f\in L^1(\BR)$ and $g\equiv 1\in L^\infty(\BR)$. Then ${\rm P}_t g\in L^\infty(\BR)$
by (1) above. In the same way as in the proof of the symmetry of ${\rm U}_t$ in Theorem \ref{k5} it can be
shown that
$$
\ix {f(x)} \cdot {\rm P}_tg(x) = \ix {{\rm P}_t f(x)}\cdot g(x)= \ix {{\rm P}_t f(x)}.
$$
Since ${\rm P}_t f(x)\geq 0$, we have $\|{\rm P}_t f\|_1 \leq \|f\|_1\|{\rm P}_t 1\|_\infty$. Taking any $f\in
L^1(\BR)$ and splitting it off as $f=\Re f_+-\Re f_-+i (\Im f_+-\Im f_-)$, we get $\|{\rm P}_t \|_1\leq 4\|
f\|_1\|{\rm P}_t 1\|_\infty$. This gives (2).

Combining (1) and (2) with the Riesz-Thorin theorem we deduce that ${\rm P}_t$ is a bounded operator from
$L^p(\BR)$ to $L^p(\BR)$, for all $1\leq p\leq \infty$. Moreover, the Markov property of $(X_t)_{t\geq0}$
implies that ${\rm P}_t$ is a semigroup on $L^p(\BR)$, for $1\leq p\leq \infty$.

Finally we consider (3) with the diagram
\eq{sk}
L^1(\BR)\stackrel{{\rm P}_t}{\longrightarrow } L^2(\BR)\stackrel{{\rm P}_t}{\longrightarrow }
L^\infty (\BR).
\en
Let $f\in\LR$. Then
$$
|{\rm P}_t f(x)|^2\leq \EE^x[e^{-2\int_0^t V(X_s) ds}] \EE^x[|f(X_t)|^2] \leq C_t
\int_\BR  |f(x+y)|^2
p_t(y)dy
$$
by Lemma \ref{muyami}, where $C_t=\sup_{x\in\BR} \EE^x[e^{-\int_0^t V(X_s)ds}]$. Since
$$
|p_t(y)| \leq \int_0^\infty e^{-t\bern( u^2/2)}du < \infty
$$
by Assumption \ref{a}, with $p_t$ in (\ref{pt}), it follows that
\eq{asa}
\|{\rm P}_t f\|_\infty \leq \lk C_t\|p_t\|_\infty \rk^{\han} \|f\|_2.
\en
Thus ${\rm P}_t$ is a bounded operator from $\LR$ to $L^\infty(\BR)$. Next, let $f\in L^1(\BR)$ and
$g\in \LR$. We have $\d \ix {\rm P}_tf(x)\cdot g(x)=\ix f(x)\cdot {\rm P}_tg(x).$ Then by \kak{asa} we
obtain
$$
\left|\ix {\rm P}_tf(x)\cdot g(x)\right| \leq \|{\rm P}_t g\|_\infty \|f\|_1\leq
(C_t\|p_t\|_\infty)^\han
\|g\|_2\|f\|_1.
$$
Since $g\in \LR$ is arbitrary, ${\rm P}_t f\in \LR$ and
\eq{yoru}
\|{\rm P}_t f\|_2\leq
(C_t\|p_t\|_\infty)^\han\|f\|_1
\en
follows, hence ${\rm P}_t$ is a bounded operator from $L^1(\BR)$ to $\LR$. Thus \kak{sk} holds.

By the semigroup property and \kak{sk} we have for $f\in L^1(\BR)$,
$$
\|{\rm P}_t f\|_\infty=
\|{\rm P}_{t/2}{\rm P}_{t/2}f\|_\infty
\leq (C_{t/2}\|p_{t/2}\|_\infty)^\han
\|{\rm P}_{t/2}f\|_2
\leq C_{t/2}\|p_{t/2}\|_\infty \|f\|_1.
$$
The fact $\|e^{-tK^\bern}\|_{p,q} \leq \|e^{-tK^\bern_0}\|_{p,q}$ follows from
\kak{huhigi}.
This completes the proof of the theorem.
\qed

\section{The case of operators with spin }
\subsection{Schr\"odinger operator with spin $\han$}
Besides operators describing interactions with magnetic fields we now consider operators also including
a spin variable. The Schr\"odinger operator with spin $\han$ is formally given by
\eq{s2}
h_\han =\half(\VS  \cdot ({\rm p}-a))^2
\en
on $L^2(\RR^3;\CC^2)$, where $\VS =(\s_1,\s_2,\s_3)$ are the Pauli matrices
$$
\s_1:= \mmm 0 1 1 0, \quad \s_2:=\mmm 0 {-i} i 0 , \quad \s_3:=\mmm 1 0 0 {-1},
$$
satisfying $\{\s_\mu, \s_\nu\}=2\delta_{\mu\nu}1$ and $\s_\mu\s_\nu=i\sum_{\la =1}^3
\epsilon^{\la \mu\nu}\s_\la $, where $\epsilon^{\la \mu\nu}$ is the anti-symmetric Levi-Civit\`a
tensor with $\epsilon^{123}=1$. We use the identification $L^2(\RR^3;\CC^2)\cong \CC^2\otimes L^2(\RR^3)$.
A rigorous definition of $h_\han $ can be given through a quadratic form in the same fashion as in the
spinless case. Define the quadratic form
\eq{y100}
q_\han (f,g)=\sum_{\mu=1}^3(\s_\mu {\rm D}_\mu f, \s_\mu {\rm D}_\mu g)
\en
with domain
$$
Q(q_\han )=\{f\in L^2(\RR^3;\CC^2)\,|\,\s_\mu {\rm D}_\mu f\in L^2(\RR^3;\CC^2),\, \mu=1,2,3\}.
$$
Assume (A1); then $q_\han $ is nonnegative and closed. By this property there exists a unique
self-adjoint operator $h_\han$ satisfying
\eq{w123}
(h_\han f,g)=q_\han (f,g),\quad f\in D(h_\han ), \;  g\in Q(q_\han),
\en
where
\eq{w1234}
D(h_\han )=\left\{f\in Q(q_\han )\,|\,q_\han (f,\cdot)\in L^2(\RR^3;\CC^2)'\right\}.
\en
\bt{yy4}
The following holds on the cores of $h_\han$:
\begin{itemize}
\item[(1)]
Let Assumption (A1) hold with $d=3$. Then $\CC^2\otimes C_0^\infty(\RR^3)$ is a form core of
$h_\han$.
\item[(2)]
Let Assumption (A4) hold. Then $\CC^2\otimes C_0^\infty(\RR^3)$ is an operator core of
$h_\han$.
\end{itemize}
Let $\bern \in \BERN$. Then furthermore the following holds on the cores of $\bern(h_\han)$:
\begin{itemize}
\item[(3)]
Take Assumption (A1) with $d=3$. Then $\CC^2\otimes C_0^\infty(\RR^3)$ is a form core of
$\bern (h_\han)$.
\item[(4)]
Take Assumption (A4). Then $\CC^2\otimes C_0^\infty(\RR^3)$ is an operator core of
$\bern (h_\han)$.
\end{itemize}
\et
\proof
The proofs of (1) and (2) are similar to that of Proposition \ref{y4}, while those of (3) and
(4) can be proven in the same way as in Theorem \ref{main2}.
\qed

Note that under Assumption (A4)
\eq{exp}
h_\han f=\half {\rm p}^2  f -a\cdot {\rm p} f + \lk -\half a\cdot a -({\rm p}\cdot a)-
\half \s \cdot (\nabla\times a)\rk f
\en
holds for $f\in \CC^2\otimes C_0^\infty(\RR^3)$. In order to construct a functional integral
representation for $e^{-th_{1/2}}$ we make a unitary transform of the operator $h_{1/2}$ on
$L^2({\mathbb  R}^3; {\mathbb  C}^2)$ to an operator on the space $L^2({\mathbb  R}^3 \times
{\mathbb  Z}_2)$. This is a space of $L^2$-functions of $x \in {\mathbb  R}^3$ and an additional
two-valued spin variable $\c \in {\mathbb  Z}_2$, where
\eq{ko25}
\mathbb Z_2= \{-1,1\}=\{\c_1,\c_2\}.
\en
Also, we define on $L^2(\RR^3\times \mathbb Z_2 )$ the operator
\begin{eqnarray}
&&
(h_{\mathbb Z_2}   f)(x,\c):=
(hf)(x,\c)-\half \c b_3(x)  f(x,\c)
- \half \lk \skima b_1(x)-i\c  b_2(x) \rk f(x,-\c),\ \ \ \ \ \
\label{yasumi}
\end{eqnarray}
where $x\in\RR^3$,  $\c\in\mathbb Z_2$ and \eq{rot} (b_1,b_2,b_3)=\nabla\times a.
\en
The closure of ${h_{\mathbb Z_2} \lceil_{\CC^2\otimes C_0^\infty(\RR^3)}}$ will be denoted by the
same symbol $h_{\mathbb Z_2} $. Also, we use the identification $L^2(\RR^3\times \mathbb Z_2) \cong
\ell^2(\mathbb Z_2)\otimes L^2(\RR^3)$. The operators $h_{\mathbb Z_2}$ and $h_\han$ are unitary
equivalent, as seen below. Define the unitary operator
\eq{uni1}
\U:L^2(\RR^3\times \mathbb Z_2 )\rightarrow \CC^2 \otimes L^2(\RR^3) \cong
L^2(\RR^3)\oplus L^2(\RR^3)
\en
by
\eq{uni2}
\U f=\vvv {f(\cdot,+1)\\ f(\cdot,-1)},\quad f\in L^2(\RR^3\times \mathbb Z_2).
\en
\bp{ko10}
Under Assumption (A4),  $h_{\mathbb Z_2}  $ is self-adjoint on $\ell^2(\mathbb Z_2)\otimes D(h)$ and
essentially self-adjoint on $\ell^2(\mathbb Z_2)\otimes C_0^\infty(\RR^3)$. Moreover,
it follows that
\eq{jin}
\U h_\han \U^{-1}=h_{\mathbb Z_2} .
\en
\ep
\proof
It can be directly seen that $\U h_\han  \U^{-1}=h_{\mathbb Z_2} $ holds on $\ell^2(\mathbb Z_2)\otimes
C_0^\infty(\RR^3)$ and $\U $ maps $\ell^2(\mathbb Z_2)\otimes C_0^\infty(\RR^3)$ onto
$\CC^2
\otimes C_0^\infty(\RR^3)$. Moreover, $\CC^2\otimes C_0^\infty(\RR^3)$ is a core of $h_\han$ by Theorem
\ref{yy4}, which yields the proposition.
\qed

\subsection{Generalized Schr\"odinger operator with spin $\mathbb Z_p$, $p\geq 2$}
Next we generalize $h_{\mathbb Z_2}$ on $L^2(\RR^3\times \mathbb Z_2)$ to consider an operator on
$L^2(\BR\times \mathbb Z_p)$ for $d\geq 1$ and $p\geq 2$. Define $\mathbb Z_p$ as the cyclic
group of the $p$th roots of unity by
\eq{ih1}
\mathbb Z_p=\{ \spin_1^{(p)},...,\spin_p^{(p)}\},
\en
where
\eq{ih123}
\spin_\alpha^{(p)}=\exp\lk 2\pi i \frac{\alpha}{p}\rk,\quad \alpha\in \mathbb N.
\en
In what follows we fix $p\geq2$
and abbreviate $\spin^{(p)}_\beta $ simply to $\spin_\beta$ for notational convenience. Consider the
finite dimensional vector space
$
\ell^2(\mathbb Z_p)=\{f:\mathbb Z_p \rightarrow \mathbb C\}
$
equipped with the scalar product
$
(f,g)_{\ell^2(\mathbb Z_p)}=\sum_{\beta=1}^p \ov{f(\spin_\beta)} g(\spin_\beta).
$

Now we consider the Schr\"odinger operator with spin $\mathbb Z_p$. We define a spin operator with its
diagonal part $U$ and off-diagonal part $U_\beta$, $\beta=1,...,p-1$, separately.
\begin{definition}[\textbf{Generalized spin operator}]
{\rm
We define two functions below:
\label{yaru}
\begin{itemize}
\item[(1)]{{(Diagonal part)}\;}
Let $U: \BR \times \mathbb Z_p \rightarrow \RR$ be such that $\max_{\spin\in
\mathbb Z_p}|U(x, \spin)|$ is a multiplication operator, relatively bounded with
respect to $\halfp$.
\item[(2)]{{(Off-diagonal part)}\;}
Let $W_\beta:\BR \times \mathbb Z_p \rightarrow \mathbb C$, $1\leq \beta\leq p-1$,
be such that $\max_{\spin\in  \mathbb Z_p}|W_\beta (x,\spin)|$ is a
multiplication operator, relatively bounded with respect to $\halfp $. Moreover, let
$U_\beta:\BR\times \mathbb Z_p\rightarrow \mathbb C$ be defined
\eq{tb2}
U_\beta(x,\spin_\alpha) = \half \lk {W_\beta(x,\spin_{\alpha+\beta})+
\ov{W_{p-\beta}(x,\spin_\alpha)}}\rk, \quad \alpha=1,...,p, \ \beta=1,...,p-1.
\en
\end{itemize}
\vspace{-0.3cm}
Furthermore, we call $\M:L^2(\BR\times\mathbb Z_p)\rightarrow L^2(\BR\times\mathbb Z_p)$,
\eq{ko26}
\M:f(x,\spin_\alpha)\mapsto U(x,\spin_\alpha) f(x,\spin_\alpha)+\sum_{\beta=1}^{p-1}
U_\beta(x,\spin_\alpha)f(x,\spin_{\alpha+\beta})
\en
\emph{generalized spin operator} on $L^2(\BR\times\mathbb Z_p)$.
}
\end{definition}

Below we will use the notation
\eq{ko53}
u_\beta(x)=\lkk
\begin{array}{ll}
\max_{\spin\in\mathbb Z_p}|U(x,\spin)| & \;\mbox{if}\;\; \beta=p,\\
& \\
\max_{\spin\in\mathbb Z_p} |U_\beta(x,\spin)|
& \;\mbox{if}\;\; 1\leq \beta\leq p-1.
\end{array}
\right.
\en
Clearly, $u_\beta(x)$ is a multiplication operator relatively bounded with respect to
$\halfp$, i.e.,
there exist $c_\beta >0$ and  $b_\beta\geq 0$ such that
\eq{rel}
\|u_\beta f\|\leq c_\beta \|\half {\rm p}^2  f\|+b_\beta \|f\|,\quad \beta=1,...,p,\en
for  all $f\in D((\han){\rm p}^2 )$. These definitions of $U$ and $U_\beta$ cover, in
particular, the $\mathbb Z_2$ case of the Schr\"odinger operator associated with spin $\han$.
\begin{example}{\bf (Spin 1/2)}
\rm{
\label{sirai}
Let $d=3$ and $p=2$. Define
$$
W_1(x,\c)=-\half (b_1(x)+i\c b_2(x)),\quad \c\in \mathbb Z_2.
$$
Then $\c_1=-1$, $\c_2=1$ and by \kak{tb2} we see that
$$
U_1(x,\c)=\half(W_1(x,\c\c_1)+\ov{W_1(x,\c)}),\quad \c\in\mathbb Z_2.
$$
It is straightforward to see that $W_{1}(x,\c\c_1)=-\half (b_1(x)-i\c b_2(x))=\ov{W_{1}(x,\c)}$,
hence the off-diagonal part is $U_1(x,\c)=-\half (b_1(x)-i\c  b_2(x))$, while the diagonal
part is given by $U(x,\c)=-\half \c b_3(x)$, both of which coincide with the interaction in
\kak{yasumi}
}
\end{example}
\begin{example}
\rm{
Let $p\geq 2$, and $W_\beta(\spin)=W(\spin)=-\half(b_1+i\spin b_2)$ for $1\leq \beta\leq p-1$. Then
\eq{hitori}
U_\beta(\spin_\alpha)=\half \left(W_\beta(\spin_{\alpha+\beta})+\ov{W_{p-\beta}(\spin_\alpha)}\right)
=-\half\lk b_1+i\frac{\spin_{\alpha+\beta}-{\spin_{p-\alpha}}}{2}b_2\rk.
\en
This gives one possible generalization of the case of spin 1/2 of Example \ref{sirai}.
}
\end{example}


\begin{definition}[\textbf{Schr\"odinger operator with generalized spin}]
{\rm
Let $h$ be the generalized Schr\"odinger operator defined in \kak{y123}. Under Assumption
(A1) we define the \emph{Schr\"odinger operator with generalized spin} $\M$ by
\eq{main55}
\hgs=1\otimes h+\M.
\en
}
\end{definition}
Above we made the identification $L^2(\BR\times \mathbb Z_p)\cong\ell^2(\mathbb Z_p)
\otimes\LR$. Formally, $\hgs$ is written as
\eq{ko27}
(\hgs f)(x,\spin_\alpha)=\lk \half ({\rm p}-a(x))^2 + U(x,\spin_\alpha) \rk
f(x, \spin_\alpha)+\sum_{\beta=1}^{p-1} U_\beta(x,\spin_\alpha)f(x,\spin_{\alpha+\beta}).
\en

\bt{self1}
Take Assumption (A2) and let $U$, $U_\beta$ be given as in Definition \ref{yaru}. Suppose
$\sum_{\beta=1}^p c_\beta<1$, where $c_\beta$ is the constant in \kak{rel}. Then
$\hgs $ is self-adjoint on $\ell^2(\mathbb Z_p)\otimes D(h)$ and bounded from below. Moreover,
it is essentially self-adjoint on any core of $1\otimes h$. In particular, $\ell^2(\mathbb Z_p)
\otimes C_0^\infty(\BR)$ is an operator core of $\hgs$.
\et
\proof
It can be seen that
$$
\sum_{\alpha=1}^p
\ov{g(x,\spin_\alpha)} \lk\sum_{\beta=1}^{p-1} {W_\beta(x,\spin_{\alpha+\beta})}
f(x,\spin_{\alpha+\beta}) \rk
=\sum_{\gamma=1}^p \lk \sum_{\beta=1}^{p-1} \ov{{\ov{W_{p-\beta}(x,\spin_{\gamma})}}
g(x,\spin_{\gamma+\beta})} \rk f(x,\spin_{\gamma})
$$
for each $x\in\BR$.
Then it follows that
$$(g(x,\cdot), \M f(x,\cdot))_{\ell^2(\mathbb Z_p)}=
(\M g(x,\cdot), f(x,\cdot))_{\ell^2(\mathbb Z_p)}$$ and $\M$ is symmetric.
Its norm can be estimated
as
$
\|\M f\| \leq
  \sum_{\beta=1}^p
\|(1\otimes u_\beta) f\|
$
by Definition \ref{yaru}. Then with $h_0=\halfp $ and $E>0$, we have by \kak{ko20} in the proof of Theorem \ref{e},
$\|u_\beta (h+E)^{-1}g\|\leq \|u_\beta (h_0+E)^{-1}|g|\|$ and hence
\begin{eqnarray*}
\|\M f\|
\leq
\sum_{\beta=1}^{p}
\| u_\beta(h_0+E)^{-1}\| \|1\otimes (h+E)f\|
\leq
\sum_{\beta=1}^p c_\beta \|(1\otimes h)  f\|+b\|f\|
\end{eqnarray*}
with a suitable constant $b$. Thus the claim follows by the Kato-Rellich theorem.
\qed

\begin{definition}
{\bf (Generalized Schr\"odinger operator with spin)}
{\rm
Suppose that $U$ and $U_\beta$ are given as in Definition \ref{yaru} and let Assumption
(A2) hold. Moreover, assume that $\sum_{\beta=1}^p c_\beta<1$. Let $\bern\in \BERN$ and put
\eq{put}
\ov{\hgs}=\left\{
\begin{array}{ll}
\hgs        &  {\rm if} \
\is{\hgs}\geq 0,\\
&  \\
\hgs-\is{\hgs}& {\rm if} \
\is{\hgs}<0 .
\end{array}
\right.
\en
We call the operator
\eq{bo15}
\HGS =\bern\left(\ov{\hgs}\right)+V
\en
\emph{generalized Schr\"odinger operator with vector potential $a$ and spin $\mathbb Z_p$}.
}
\end{definition}

\bc{ko12}
Let $U$ and $U_\beta$ be given as in Definition \ref{yaru}, assume (A2) and suppose that $\sum_{\beta=1}^p c_\beta<1$.
If $\bern\in \BERN$, then $\ell^2(\mathbb Z_p)\otimes C_0^\infty(\BR)$ is an operator core of $\bern (\hgs)$.
\ec
\proof
Since $\hgs$ is essentially self-adjoint on $\ell^2(\mathbb Z_p)\otimes C_0^\infty(\BR)$, the corollary can
be proven in the same way as Theorem \ref{main2}.
\qed

\subsection{Functional integral representation}
In this subsection we give a functional integral representation of $e^{-t\HGS}$ by means of Brownian motion,
a jump process and a subordinator.

Let $(N^\beta_t)_{t\geq 0}$, $\beta=1,...,p-1$, be $p-1$ independent Poisson processes with unit intensity
on a probability space $(\PO, \ms F_N, \mu)$, i.e., $\d \mu(N^\beta _t=n)=e^{-t}t^n/n!$. Define the random
process $(N_t)_{t\geq 0}$ by
\begin{equation}
N_t=\sum_{\beta=1}^{p-1} \beta N_t^\beta.
\end{equation}
Let $\ms F^N_t=\s(N_t,t\leq s)$ be the natural filtration. Then since $N_t$ is a L\'evy process, it is a
Markov process with respect to $\ms F^N_t$.
We write $\mathbb E _\mu[f(N_t+\alpha)]$ as $\mathbb E_\mu^\alpha[f(N_t)]$. Also, $\mathbb E_\mu^
\alpha[N_0=\alpha]=1$. Define
\eq{ko13}
\int_v^{w+} g(N_{s-}) dN_s^\beta
=\sum_{
{v\leq r\leq w}
\atop
{N_{r+}^\beta\not= N_{r-}^\beta}}
g(N_{r-}).
\en
It can be seen that
\eq{exp1}
\mathbb E_\mu\left[\int_v^{w+} g(N_{s-}) dN_s^\beta\right]=\mathbb E_\mu\left[\int_v^{w} g(N_{s}) ds\right].
\en
The next lemma is an extension of a result obtained in \cite{als83,hl08}.
\bl{Main}
Let $U$ and $U_\beta$ be given in Definition \ref{yaru} and assume $\sum_{\beta=1}^p c_\beta<1$.
Suppose
Assumption
(A2).
and
\eq{cont}
\d \int_0^t ds \int_\BR dy (2\pi s)^{-d/2} e^{-|x-y|^2/(2s)} |\log u_\beta (y)|  <\infty,\quad \beta=1,...,p-1.
\en
Then
\eq{main 777}
(f, e^{-t\hgs}g) = e^{(p-1)t} \sum_{\alpha=1}^p \ix \mathbb E_{P\times \mu}^{x,\alpha}
\lkkk \ov{f(B_0,\spin_{N_0})}g(B_t, \spin_{N_t})e^{\SSS}\rkkk,
\en
where $\SSS = \SSS_a + \SSS_{\rm spin}$ and
\begin{eqnarray*}
&& \SSS_a=-i\int_0^t a(B_s)\circ dB_s,\\
&& \SSS_{\rm spin} =-\int_0^t  U(B_s,\spin_{N_s}) ds+ \sum_{\beta=1}^{p-1} \int_0^{t+ }
\log(-U_\beta(B_s,\spin_{N_{s-}})) dN_s^\beta.
\end{eqnarray*}
Here we take $\log z$ with the principal branch for $z\in\CC$.
\el
\proof
First assume that the diagonal part $U(x,\spin_\alpha)$ and off-diagonal part $U_\beta(x,\spin_\alpha)$
are continuous in $x$
and $a\in (C_0^\infty(\BR))^d$.
Since from \kak{cont} and
\kak{exp1} it follows that
$$
\mathbb E_{P\times\mu}^{x,\alpha} \lkkk \int_0^{t+} |\log(-U_\beta(B_s, \spin_{N_{s-}}))| dN_s^\beta \rkkk
\leq \int_0^t ds \int_\BR \frac{e^{-|x-y|^2/(2s)}} {(2\pi s)^{d/2}} |\log u_\beta(y)|<\infty,
$$
we note that
\eq{ko15}
\int_0^{t+}|\log(-U_\beta(B_s, \spin_{N_{s-}}))| dN_s^\beta<\infty
\en
almost surely. By the estimate $|c\SSS_{\rm spin}| \leq c \|u_p\|_\infty t +|\log \|u_\beta\|_\infty^c|N_t^\beta$
and the equality
$$
\int_{\Omega_0}\exp\left(\sum_{\beta=1}^{p-1} r_\beta N_t^\beta\right) d\mu =
\exp\lk \sum_{\beta=1}^{p-1}(e^{r_\beta}-1) \rk
$$
for $r_\beta\in \RR$, we have for $c>0$,
\begin{equation}
\left|\mathbb E_{P\times\mu}^{x,\alpha}[e^{c \SSS_{\rm spin}}]\right| \leq
\exp\left(t \lk c \|u_p\|_\infty +\sum_{\beta=1}^{p-1}(\|u_\beta\|_\infty ^c-1) \rk \right),
\label{cc}
\end{equation}
where $u_\beta$ is given in \kak{ko53}, and $\sup_x \mathbb E_{P}^{x} \lkkk e^{4\SSS_V}\rkkk <\infty$. Denote
\begin{eqnarray*}
Z_{[v,w]} = -i\int_v^w a(B_s)\circ dB_s -\int_v^w U(B_s, \spin_{N_s}) ds
+\sum_{\beta=1}^{p-1} \int_v^{w+}\log(-U_\beta(B_s,\spin_{N_{s-}})) dN_s^\beta
\end{eqnarray*}
and let
$$
{\rm P}_tg(x,\spin_\alpha)=\mathbb E_{P\times \mu }^{x,\alpha}\lkkk e^{Z_{[0,t]}}
g(B_t, \spin_{N_t}) \rkkk.
$$
Let $g\in \ell^2(\mathbb Z_p)\otimes
C_0^\infty(\BR)$.
By the Schwarz inequality and setting $c=2$ in (\ref{cc}) we have the estimate
\begin{eqnarray*}
\|{\rm P}_t g\|^2
&\leq&
\sum_{\alpha=1}^p \ix \mathbb E_{P\times \mu}^{x,\alpha} \lkkk g(B_t, \spin_{N_t})^2\rkkk
\mathbb E_{P\times \mu}^{x,\alpha}
\lkkk e^{2\SSS_{\rm spin}}\rkkk \\
&\leq&
\exp\left(t\left(2\|u_p\|_\infty
+\sum_{\beta=1}^{p-1}(\|u_\beta\|_\infty^2-1)\right)\right) \|g\|^2.
\end{eqnarray*}
Thus ${\rm P}_t$ is bounded. We show now that $\{{\rm P}_t\}_{t\geq 0}$ is a  $C_0$-semigroup with
generator $-(\hgs+p-1)$, i.e.,
(1) ${\rm P}_0 =I$,
(2) ${\rm P}_s{\rm P}_t={\rm P}_{s+t}$,
(3) ${\rm P}_tg$ is continuous in $t$ and (4) $\d
\lim_{t\rightarrow 0}\frac{1}{t} ({\rm P}_tg-g)=-(\hgs +(p-1))g$ in strong sense. First, (1)
is trivial. To check (2) notice that
\eq{hi3}
{\rm P}_t {\rm P}_s g(x,\spin_\alpha)=\mathbb E_{P\times \mu }^{x,\alpha} \lkkk e^{Z_{[0,t]}}
\mathbb E_{P\times \mu }^{B_t, N_t} \lkkk e^{Z_{[0,s]}} g(B_s, \spin_{N_s}) \rkkk \rkkk.
\en
By the Markov property of $B_t$ we have
\begin{eqnarray}
&&
\hspace{-0.5cm}
\kak{hi3}=\mathbb E_{P\times \mu }^{x,\alpha}\! \lkkk e^{Z_{[0,t]}}
\exp\lk {-i\int_t^{t+s} a(B_r)\circ dB_r}\rk \right. \non \\
&&
\hspace{-0.5cm}
\left.
\mathbb E_{\mu }^{N_t}\!\! \lkkk \exp\lk\!\! -\!\!\int_0^s U(B_{t+r}, \spin_{N_r}) dr  +\sum_{\beta=1}^{p-1}
\int_0^{s+}\!\!\!\log(-U_\beta(B_{t+r-},\spin_{N_{r-}})) dN_r^\beta\rk g(B_{t+s}, \spin_{N_s}) \rkkk \rkkk.
\non\\
&&
\label{hi4}
\end{eqnarray}
Furthermore the Markov property of $N_t$ yields that
$$
\kak{hi4}=\mathbb E_{P\times \mu }^{x,\alpha} \lkkk e^{Z_{[0,t]}}e^{Z_{[t,t+s]}}g(B_{t+s},\spin_{N_{t+s}})
\rkkk ={\rm P}_{s+t} g(x,\spin_\alpha).
$$
This proves the semigroup property (2). Next we obtain the generator of ${\rm P}_t$. An application of the
It\^{o} formula (see Appendix A) yields that
$$\d dN_t =\sum_{\beta=1}^{p-1}\int_0^{t+}\beta dN_s^\beta,\quad
d\theta_{N_t}=\sum_{\beta=1}^{p-1} (\theta_{N_t+\beta}-\theta_{N_t})$$ and
\begin{eqnarray*}
dg(B_t,\spin_{N_t})
&=&
\int_0^t \nabla g(B_s,\spin_{N_s}) \cdot dB_s +\half \int_0^t \Delta g(B_s,\spin_{N_s}) ds \\
&&
\qquad\qquad\qquad
+\sum_{\beta=1}^{p-1}\int_0^{t+} (g(B_s, \spin_{N_s+\beta})-g(B_s, \spin_{N_s}))dN_s^\beta\\
de^{Z_{[0,t]}}
&=&
\int_0^t e^{Z_{[0,s]}} (-ia(B_s))\cdot dB_s+\half \int_0^t e^{Z_{[0,s]}}
(-i\nabla\cdot a(B_s)-a(B_s)^2) ds\\
&&
-\int_0^t e^{Z_{[0,s]}} U(B_s, \spin_{N_s})ds+\sum_{\beta=1}^{p-1} \int_0^{t+}e^{Z_{[0,s-]}}\lk
e^{\log(-U_\beta(B_s, \spin_{N_{s-}}))}-1\rk dN_s^\beta.
\end{eqnarray*}
The product formula (Appendix A) $d \lk e^{Z_{[0,t]}} g\rk  = d e^{Z_{[0,t]}} \cdot g+ e^{Z_{[0,t]}}
\cdot d g+ d e^{Z_{[0,t]}} \cdot d g$ furthermore gives
\begin{eqnarray*}
\lefteqn{
d \lk e^{Z_{[0,t]}} g\rk (B_t, \spin_{N_t})
=\int_0^t e^{Z_{[0,s]}} \lkk \half \Delta g(B_s,\spin_{N_s})-ia(B_s)\cdot \nabla g(B_s,\spin_{N_s})\right.}\\
&&
\hspace{6cm}
\left.
+ \lk -\half a(B_s)^2-U(B_s,\spin_{N_s})\rk g(B_s,\spin_{N_s})\rkk ds\\
&&
\hspace{2cm}
+\int_0^t e^{Z_{[0,s]}} \lk \skima \nabla g(B_s,\spin_{N_s}) -ia(B_s) g(B_s,\spin_{N_s}) \rk \cdot dB_s\\
&&
\hspace{2cm}
+ \sum_{\beta=1}^{p-1} \int_0^{t+} e^{Z_{[0,s]}} \lk g(B_s,\spin_{N_{s-}+\beta})
e^{\log(-U_\beta (B_s,\spin_{N_{s-}}))}-g(B_s,\spin_{N_{s-}}) \rk dN_s^\beta.
\end{eqnarray*}
Taking expectation values on both sides above yields
\begin{eqnarray*}
\frac{1}{t}(f, ({\rm P}_t-1)g)=\frac{1}{t}\int_0^t ds \int_\BR dx \ov{f(x)}
\mathbb E_{P\times \mu}^{x,\alpha}\lkkk G(s)\rkkk,
\end{eqnarray*}
where
\begin{eqnarray*}
&&
G(s)= e^{Z_{[0,s]}} \lk \half \Delta-ia(B_s)\cdot \nabla -\half a(B_s)^2-U(B_s,\spin_{N_s})\rk
g(B_s,\spin_{N_s})\\
&&
\hspace{2cm}
+\sum_{\beta=1}^{p-1} e^{Z_{[0,s]}} \lk g(B_s,\spin_{N_s+\beta})e^{\log(-U_\beta (B_s,\spin_{N_s}))}
-g(B_s,\spin_{N_{s-}}) \rk,\\
&&
G(0)= \lk \half \Delta-ia(B_0)\cdot \nabla-\half a(B_0)^2-U(B_0,\spin_{N_0})\rk g(B_0,\spin_{N_0})\\
&&
\hspace{2cm}
+\sum_{\beta=1}^{p-1} \lk -U_\beta (B_0,\spin_{N_0}) g(B_0,\spin_{N_0+\beta})-g(B_0,\spin_{N_{0}})\rk\\
&&
\quad \quad \quad
=-(\hgs +(p-1))g(x,\spin_\alpha).
\end{eqnarray*}
Note that $U(x,\c)$, $U_\beta(x,\c)$, $a_\mu(x)$ are continuous in $x$. Therefore $G(s)$ is
continuous at $s=0$ for each $(\omega,\tau)\in \BM\times \PO$, and $\mathbb E_{P\times \mu}^
{x,\alpha}[G(s)]$ is continuous at $s=0$ by the dominated convergence theorem. Thus
$$
\lim_{t\rightarrow 0}\frac{1}{t}(f, ({\rm P}_t-1)g)=(f, -(\hgs +(p-1))g)
$$
follows.
Finally, the strong continuity (3)
follows from (2) and (4),
and hence
\eq{ko30}
e^{t(p-1)} {\rm P}_t g=e^{-t\hgs }g.
\en
By a similar approximation argument as in the proof of Proposition \ref{y77},
\kak{ko30} can be extended to $a$ obeying Assumption (A2). 
Finally, we extend
\kak{ko30} for $U$ and $U_\beta$ given in Definition \ref{yaru}. By using a mollifier it is seen that there
exists a sequence $U_\beta^{(n)}(x,\spin_\alpha)$ and $U^{(n)}(x,\spin_\alpha)$, $n=1,2,3,...$, such that
they are continuous in $x$ and converge to $U_\beta(x,\spin_\alpha)$ resp. $U(x,\spin_\alpha)$ for each $x$
as $n\to\infty$, and $\|U^{(n)}(\cdot,\spin_\alpha)\|_\infty \leq \| U(\cdot,\spin_\alpha)\|_\infty$ and
$\|U_\beta^{(n)}(\cdot,\spin_\alpha)\|_\infty \leq \|U_\beta(\cdot,\spin_\alpha)\|_\infty$. For each fixed
$\tau\in \PO$ there exists $r_1=r_1(\tau),...,r_M=r_M(\tau)$, where $M=M(\tau)$, such that
\eq{ko31}
\exp\left(\sum_{\beta=1}^{p-1} \int_0^{t+} \log(-U_\beta(B_s, \spin_{N_{s-}}) )dN_s^\beta\right)=
\prod_{\beta=1}^{p-1} \prod_{i=1}^{M} (-U_\beta(B_{r_i},\spin_{N_{r_i}})).
\en
Then for each $\tau\in \PO$,
\begin{eqnarray}
\lefteqn{
\hspace{-2cm}
\lim_{n\rightarrow \infty} \exp\left(\sum_{\beta=1}^{p-1}
\int_0^{t+}\log(-U^{(n)}_\beta(B_s, \spin_{N_{s-}}) )dN_s^\beta\right) \non} \\
&&
\label{ko32}
= \exp\left(\sum_{\beta=1}^{p-1}\int_0^{t+} \log(-U_\beta(B_s, \spin_{N_{s-}}) )dN_s^\beta\right).
\end{eqnarray}
In the same way as above we can also see that $e^{-\int_0^t U^{(n)}(B_s, \spin_{N_s}) ds} \rightarrow
e^{-\int_0^t U(B_s,\spin_{N_s}) ds}$ as $n\rightarrow \infty$ almost surely. Therefore by the dominated
convergence theorem \kak{ko30} holds for such $U_\beta$ and $U$. \qed

Now we can state and prove the functional integral representation of $e^{-t\hgss}$.
\bt{MAIN}
Let $\bern\in\BERN$, and  $U$, $U_\beta$ be given as in Definition \ref{yaru}. Assume $u_\beta
\in L^\infty(\BR)$, $\beta=1,...,p$, $V\in L^\infty(\BR)$, and let Assumption (A2) and
\eq{cont22}
 \int_\RR \rho(r,t) dr \int_0^r ds \int_\BR dy (2\pi s)^{-d/2} e^{-|x-y|^2/(2s)} |\log u_\beta (y)|  <\infty,\quad \beta=1,...,p-1,
\en
where $\rho(r,t)$ is the distribution of $T_t^\bern$ on $\RR$.
Then
\eq{main 77}
(f, e^{-t \hgss} g) = \sum_{\alpha=1}^p \ix  \mathbb E_{P\times \mu\times \nu}^{x,\alpha,0}
\lkkk  e^{(p-1)T_t^\bern} \ov{f(B_0,\spin_{N_0})} g(B_{T_t^\bern }, \spin_{N_{T_t^\bern }})
e^{{\SSS}^\bern}\rkkk,
\en
where
$\SSS^\bern = \SSS_V^\bern  + \SSS_a^\bern  + \SSS_{\rm spin}^\bern $
and
\begin{eqnarray*}
&&
\hspace{-0.5cm}
\SSS_V^\bern =-\int_0^t V(B_{T_s^\bern}) ds, \\
&&\hspace{-0.5cm}
  \SSS_a^\bern =-i\int_0^{T_t^\bern } a(B_s)\circ dB_s,\\
&&\hspace{-0.5cm}
 \SSS_{\rm spin}^\bern =\left\{\begin{array}{l}
\d   -\int_0^{T_t^\bern }\lk\skima U(B_s,\spin_{N_s})-\is{\hgs} \rk ds +
\sum_{\beta=1}^{p-1} \int_0^{T_t^\bern+}\log(-U_\beta(B_s, \spin_{N_{s-}})) dN_s^\beta
  \\
 \hspace{10cm}
 {\rm if}\ \is{\hgs}<0,\\
\d
 -\int_0^{T_t^\bern }\lk\skima U(B_s,\spin_{N_s}) \rk ds +
\sum_{\beta=1}^{p-1} \int_0^{T_t^\bern+}\log(-U_\beta(B_s, \spin_{N_{s-}})) dN_s^\beta \\
\hspace{10cm}
 {\rm if}\ \is{\hgs}\geq0.
\end{array}
\right.
\end{eqnarray*}
\et
\proof
Since from \kak{cont22}
it follows that
\eqn
&&
\mathbb E_{P\times\mu\times\nu}^{x,\alpha,0} \lkkk \int_0^{T_t^\bern+} |\log(-U_\beta(B_s, \spin_{N_{s-}}))| dN_s^\beta \rkkk\\
&&
\leq \int_\RR\rho(r,t) dr
\int_0^r ds \int_\BR \frac{e^{-|x-y|^2/(2s)}} {(2\pi s)^{d/2}} |\log u_\beta(y)|<\infty,
\enn
we notice  that
\eq{ko155}
\int_0^{T_t^\bern+}|\log(-U_\beta(B_s, \spin_{N_{s-}}))| dN_s^\beta<\infty
\en
almost surely.  Using Lemma \ref{Main} we obtain
\eq{main 7}
\lk
f, e^{-t\bern(\ov{\hgs} )}g\rk = \sum_{\alpha=1}^p \ix \mathbb E_{P\times \mu\times \nu}^{x,\alpha,0}
\lkkk e^{(p-1)T_t^\bern}\ov{f(B_0,\spin_{N_0})} g(B_{T_t^\bern }, \spin_{N_{T_t^\bern }})
e^{\SSS_a^\bern +\SSS_{\rm spin}^\bern }\rkkk.
\en
Let $0=t_0<t_1<\cdots<t_n=t$. We show that
\begin{eqnarray}
\lefteqn{
\lk f_0, \prod_{j=1}^n e^{-(t_j-t_{j-1}) \bern(\ov{\hgs}) }f_j \rk  } \non\\
&&
\label{ko33}
=\sum_{\alpha=1}^p \ix \mathbb E_{P\times\mu\times\nu}^{x,\alpha,0} \lkkk e^{(p-1) T_t^\bern}
\overline{f(B_0,\spin_{N_0})} \lk
\prod_{j=1}^n f_j(B_{T^\bern_{t_j}},\spin_{N_{T^\bern_{t_j}}})
\rk
e^{\SSS_a^\bern +\SSS_{\rm spin}^\bern } \rkkk.\ \ \ \
\end{eqnarray}
This can be proven in the same way as in Step 2 of the proof of Theorem \ref{bo8} with
the $d$-dimensional Brownian motion $B_t$ on $(\BM, {\ms F_P}, P^x)$ replaced by the $d+1$
dimensional Markov process $(B_t,{N_t})$ on $(\BM\times\PO,{\ms F_P}\times
{\ms F_N},P^x\times\mu)$ under the natural filtration. Suppose $V$ is continuous. By the Trotter
product formula and \kak{ko33} it is seen that
\begin{eqnarray*}
\lk f, e^{-t\HGS}g \rk
&=&\lim_{n\rightarrow \infty} \lk f,\lk e^{-\lk t/n\rk \bern \lk \hgs\rk }e^{-\lk t/n\rk  V}\rk ^n g\rk \\
&=&
\lim_{n\rightarrow \infty} \sum_{\alpha=1}^p \ix \mathbb E_{P\times\mu\times\nu}^{x,\alpha,0} \\
&&\hspace{1cm}
\lkkk e^{\lk p-1\rk T_t^\bern} \overline{f\lk B_0,\spin_{N_0}\rk }
e^{-\sum_{j=1}^n \frac{t}{n}V\lk B_{T^\bern_{jt/n}}\rk } g\lk B_{T_t},\spin_{N_{T_t^\bern}}\rk
e^{\SSS_a^\bern +\SSS_{\rm spin}^\bern }\rkkk\\
&=&
\sum_{\alpha=1}^p \ix \mathbb E_{P\times\mu\times\nu}^{x,\alpha,0}
\lkkk e^{\lk p-1\rk T_t^\bern} \overline{f\lk B_0,\spin_{N_0}\rk } g\lk B_{T_t^\bern},\spin_{N_{T_t^\bern}}\rk
e^{\SSS^\bern} \rkkk.
\end{eqnarray*}
Hence the theorem follows for continuous $V$. This can be extended for $V\in L^\infty(\BR)$ in the same way as
in Step 4 of the proof of Theorem \ref{3t}.
\qed
In the case of $\bern(u)=\sqrt{2u+m^2}-m$, the distribution of $T_t^\bern$ is exactly given by
\kak{enta}.
\begin{remark}
{\rm
Notice that conditions \kak{cont} and \kak{cont22} depend on $t$.
Let us replace \kak{cont} and \kak{cont22}  with
the condition
\eq{cont23}
   \int_0^\infty ds \int_\BR dy (2\pi s)^{-d/2} e^{-|x-y|^2/(2s)} |\log u_\beta (y)|  <\infty,\quad \beta=1,...,p-1.
\en
Then we see that
$$
\mathbb E_{P\times\mu\times\nu}^{x,\alpha,0} \lkkk \int_0^{T_t^\bern+} |\log(-U_\beta(B_s, \spin_{N_{s-}}))| dN_s^\beta \rkkk
\leq\int_0^\infty ds \int_\BR \frac{e^{-|x-y|^2/(2s)}} {(2\pi s)^{d/2}} |\log u_\beta(y)|
<\infty
$$
and
\eqn
\mathbb E_{P\times\mu}^{x,\alpha} \lkkk \int_0^{t+} |\log(-U_\beta(B_s, \spin_{N_{s-}}))| dN_s^\beta \rkkk
\leq\int_0^\infty ds \int_\BR \frac{e^{-|x-y|^2/(2s)}} {(2\pi s)^{d/2}} |\log u_\beta(y)|
<\infty.
\enn
In particular
\eq{ko1555}
\int_0^{T_t^\bern+}|\log(-U_\beta(B_s, \spin_{N_{s-}}))| dN_s^\beta<\infty
\en
and
\eq{ko15556}
\int_0^{t+}|\log(-U_\beta(B_s, \spin_{N_{s-}}))| dN_s^\beta<\infty
\en
follow for {\it all} $t\geq0$.
}\end{remark}

Now let $\hgs^0$ be defined by $\hgs$ in \kak{main55} with $a$ and $U_\beta$, $\beta=1,...,p-1$, replaced by $0$
and $|U_\beta|$, respectively, i.e.,
\eq{uni3}
(\hgs^0 f)\lk x, \spin_\alpha\rk =\halfp  f\lk x, \spin_\alpha\rk +U\lk x, \spin_\alpha\rk  f\lk x,
\spin_\alpha\rk -\sum_{\beta=1}^{p-1}|U_\beta\lk x, \s\rk |f\lk x, \spin_{\alpha+\beta}\rk .
\en
Let
\eq{imm}
\ov{\hgs^0}=\lkk
\begin{array}{ll}
\hgs^0 & {\rm if}\ \is{\hgs^0}\geq0, \\ & \\
\hgs^0-\is{\hgs^0}& {\rm if}\ \is{\hgs^0}<0.
\end{array}
\right.
\en
An immediate corollary of Theorem \ref{MAIN} is
\bc{wr4}{\rm (\textbf{Diamagnetic inequality})}
Under the assumptions of Theorem \ref{MAIN} we have
\eq{bern20}
\hgs-\is{\hgs^0}\geq0.
\en
Moreover,
\begin{enumerate}
\item[(1)]
if $\is {\hgs^0}\geq 0$, then
\eq{bo1}
\left|\lk f, e^{-t\lk \bern\lk {\hgs}\rk +V\rk }g \rk \right| \leq
\lk |f|, e^{-t\lk \bern\lk {\ov{\hgs^0}}\rk +V\rk }|g| \rk
\en
and
\eq{di11}
\is{\bern\lk {\ov{\hgs^0}}\rk +V} \leq \is{\bern\lk {\hgs} \rk +V};
\en
\item[(2)]
if $\is{\hgs^0}<0$, then
\eq{bo}
\left|\lk f, e^{-t\lk \bern\lk {\hgs}-\is{\hgs^0} \rk +V\rk }g \rk\right|
\leq \lk |f|,e^{-t\lk \bern\lk \ov{\hgs^0}\rk +V\rk }|g| \rk
\en
and
\eq{di1}
\is{\bern\lk \ov{\hgs^0}\rk +V} \leq \is{\bern\lk {\hgs}-\is{\hgs^0} \rk +V}.
\en
\end{enumerate}
\ec
\proof
Note the estimate
\eq{yopp}
\left|\exp\lk \sum_{\beta=1}^{p-1} \int_0^{T^\bern_t+} \log \lk -U_\beta
\lk \spin_{N_{s-}^\beta}\rk \rk dN_s^\beta \rk \right| \leq
\exp \lk \sum_{\beta=1}^{p-1} \int_0^{T^\bern_t+} \log |U_\beta\lk \spin_{N_{s-}^\beta} \rk|
dN_s^\beta \rk.
\en
Let $\bern(u)=u$ and then $T^\bern_t=t$. Theorem \ref{MAIN} and \kak{yopp} imply that
\eq{boob}
|\lk f, e^{-t{\hgs}}g\rk | \leq \lk |f|, e^{-t{\hgs^0}}|g|\rk.
\en
This further implies $\is{\hgs^0}\leq\is{\hgs}$, thus \kak{bern20} holds. \kak{bo1} and \kak{bo} follow
similarly by Theorem \ref{MAIN} and the estimate \kak{yopp}. \kak{di11} and \kak{di1} are an immediate
consequence of \kak{bo1} and \kak{bo}, respectively.
\qed

\bt{ee}
Let $U$ and $U_\beta$ be given by Definition \ref{yaru} and suppose that $\sum_{\beta=1}^p c_\beta<1$. Let
Assumption (A2) and \kak{cont} hold, and suppose that $|V|$ is relatively bounded with respect to
$\bern\lk \ov{\hgs^0}\rk$ with a relative bound $b$. Then $|V|$ is relatively bounded with respect to
$\bern\lk {\ov{\hgs}}\rk $ with a relative bound not larger than $b$.
\et
\proof
We prove the theorem in the case of $\is{\hgs^0}<0$, the case $\is{\hgs^0}\geq 0$ is simpler. By the
assumption we have for every $\epsilon>0$,
\eq{ko40}
\| Vf\|\leq (b+\epsilon)\|\bern\lk \ov{\hgs^0}\rk f\| + c\|f\|.
\en
By virtue of Corollary \ref{wr4} we have
\eq{ko37}
\frac{\||V|\lk \bern\lk {\hgs}-\is{\hgs^0}\rk +E\rk ^{-1}f\|} {\|f\|} \leq
\frac{\||V| \lk \bern\lk \ov{\hgs^0}\rk +E\rk ^{-1}|f|\|}{\|f\|}
\en
By \kak{ko40} the right hand side of \kak{ko37} converges to a number smaller than $b+\epsilon$ as
$E\rightarrow\infty$. Thus
\eq{genki2}
\|Vf\|\leq (b+\epsilon) \|\bern\lk \hgs-\is{\hgs^0}\rk f\|+c_b\|f\|
\en
follows with some constant $c_b$. Let $X<Y$ and $X<0$. From \kak{6} we can see that
$$
\bern(u-X)-\bern(u-Y)=b(Y-X)+\int_0^\infty e^{-(u-Y)y}(1-e^{-(Y-X)y}) \la(dy),\quad u\geq Y.
$$
Hence $\sup_{u\geq Y}|\bern(u-X)-\bern(u-Y)|\leq \bern(Y-X)$. From this and $\is{\hgs^0}\leq \is{\hgs}$ we
obtain that
$$
\sup_{u\geq \is{\hgs}} |\bern(u-\is{\hgs^0})-\bern(u-\is{\hgs})| \leq \bern(\is{\hgs}-\is{\hgs^0}).
$$
Thus the spectral decomposition yields that
$$
\|\bern(\hgs-\is{\hgs^0})f\| \leq \|\bern(\hgs-\is{\hgs})f\| + \bern(\is{\hgs}-\is{\hgs^0})\|f\|.
$$
Then the theorem follows together with \kak{genki2}, since $\epsilon$ is arbitrary.
\qed
We have the immediate consequences below.
\bt{ko39}
Let $U$ and $U_\beta$ be given in Definition \ref{yaru} and assume $\sum_{\beta=1}^p c_\beta<1$.
Suppose that $V$ is relatively bounded with respect to $\bern(\ov{\hgs^0})$ with a relative bound strictly
less than 1. Moreover, assume
\kak{cont}.
\begin{itemize}
\item[(1)]
Let Assumption (A2) hold. Then $\hgss$ is self-adjoint on $D\lk \bern \lk \ov{\hgs}\rk \rk $ and
essentially self-adjoint on any core of $\bern \lk \ov{\hgs}\rk $. In particular, under Assumption (A3) the
operator $\hgss$ is essentially self-adjoint on $C_0^\infty(\BR)$.
\item[(2)]
Let Assumption (A3) hold. Then the functional integral representation of $e^{-t\hgss}$ is given by (\ref{MAIN}).
\end{itemize}
\et
\proof
(1) is trivial. (2) Let $V=V_+-V_-$. Note that $V_+, V_-$ are relatively bounded with respect to $\bernz$ with
a relative bound strictly less than 1. Define $V_{+,n}(x)=\phi(x/n)(V_+\ast j_n)$ and $V_{-,m}= \phi(x/n)
(V_-\ast j_m)$, where $\phi$ and $j_n$ are defined in Step 4 of the proof of Theorem \ref{bo8}. Notice that
$e^{-t\lk \bern\lk \ov{\hgs}\rk +V_{+,n}-V_{-,m}\rk }$ strongly converges to $e^{-t\lk \bern\lk \ov{\hgs}\rk +V\rk}$
as $n,m\rightarrow\infty$, since $\bern\lk \ov{\hgs}\rk +V_{+,n}-V_{-,m}$ converges to $\bern\lk \ov{\hgs}\rk +V$ on
the common core $\ell^2(\mathbb Z_2)\otimes C_0^\infty( \RR^3)$. Then the theorem can be proven in a similar way to
Step 4 of Theorem \ref{bo8}.
\qed

\section{Relativistic Schr\"odinger operators}
\subsection{Case of spin $\han$}
In this subsection we further discuss the functional integral representation for the specific
case of the relativistic Schr\"odinger operator with spin $\han$.
Throughout this section $d=3$ and $p=2$. Therefore, $\c_\alpha=\c_\alpha^{(2)}$, $\alpha=1,2$,
and $\c_1=-1$ and $\c_2=+1$.
The relativistic Schr\"odinger operator with spin $\han$ is given by
\eq{haikei}
h_\han^{\rm rel} =
\sqrt{2h_\han+m^2}-m, \quad m \geq 0,
\en
on $L^2(\RR^3;\CC^2)$, where
$h_\han=(\VS  \cdot ( {\rm p}-a)) ^2$.
\bigskip

\noindent
\emph{Functional integral representation. \quad}
Let Assumption (A4) hold. Then $h_\han^{\rm rel}$ is unitary equivalent to
\eq{ko34}
h_{\mathbb Z_2} ^{\rm rel}=\sqrt{2h_{\mathbb Z_2} +m^2}-m,
\en
where $h_{\mathbb Z_2} $ is defined on $L^2(\RR^3\times\mathbb Z_2)$ and given in
(\ref{yasumi}) as
$$
(h_{\mathbb Z_2}   f)(x,\c):=
\lk \half({\rm p}-a)^2f\rk (x,\c)-\half \c b_3(x) f(x,\c)
- \half \lk \skima b_1(x)-i\c  b_2(x) \rk f(x,-\c)$$
for $x\in\RR^3$ and $\c\in\mathbb Z_2$. Recall that here $b=(b_1,b_2,b_3)=\nabla\times  a$. Clearly,
$h_{\mathbb Z_2} ^{\rm rel}$ is non-negative and $h_{\mathbb Z_2} ^{\rm rel} = \bern\lk
h_{\mathbb Z_2}\rk $ with the Bernstein function $\bern(u)=\sqrt{2u+m^2}-m$. The spin operator in
$h_{\mathbb Z_2} $ is furthermore given by
\begin{eqnarray*}
\begin{array}{ll}
{\rm (diagonal\ component)} & \d U(x,\c)=-\half \c b_3(x),\\ \\
{\rm (off-diagonal\ component)} & \d U_1(x,\c)=-\half (b_1(x)-i\c b_2(x)).
\end{array}
\end{eqnarray*}
Let $h_{\mathbb Z_2}^0$ be defined by $h_{\mathbb Z_2} $ with vector potential $a\in (L_{\rm loc}^4(\RR^3))^3$
and off-diagonal component $U_1$ replaced by $0$ and $|U_1|=\half\sqrt{b_1^2(x)+b_2^2(x)}$, respectively, i.e.,
$$
(h_{\mathbb Z_2}^0 f)(x,\c)=\lk \halfp f\rk (x,\c) -\half \c  b_3(x) f(x,\c )- \half \sqrt{b_1(x)^2+b_2(x)^2}f(x,-\c).
$$
The operator $h_{\mathbb Z_2}^0$ is unitary equivalent with $h_\han^{0}$ on $L^2(\RR^3;\CC^2)$ given
by
\eq{kawa}
h_\han^{0}=\half{\rm p}^2-\half \mmm {b_3}{\sqrt{b_1^2+b_2^2}}{\sqrt{b_1^2+b_2^2}}{-b_3}.
\en
We write
\begin{eqnarray*}
&&
h_{\mathbb Z_2}^{\rm rel}(0)=\sqrt{2 \ov{h_{\mathbb Z_2}^0} +m^2}-m,\\
&&
h_\han^{\rm rel}(0)=\sqrt{2 \ov{h_\han^0} +m^2}-m,
\end{eqnarray*}
where $\ov{h_{\mathbb Z_2}^0}={h_{\mathbb Z_2}^0}-\is{{h_{\mathbb Z_2}^0}}$ and $\ov{h_\han^0} =
{h_\han^0}-\is{{h_\han^0}}$. The operators $h_\han^{\rm rel}$ and $h_{\mathbb Z_2} ^{\rm rel}$ are essentially
self-adjoint on $\CC^2\otimes C_0^\infty(\RR^3)$ and $\ell^2(\mathbb Z_2)\otimes C_0^\infty(\RR^3)$, respectively.
\bt{ko50}
Let Assumption (A4) hold and further assume (1)-(4) below:
\begin{enumerate}
\item[(1)]
$V$ is relatively bounded with respect to $\sqrt{{\rm p}^2+m^2}$ with a relative bound $A<1$;
\item[(2)]
each $-\half b_j$, $j=1,2,3$, is relatively bounded with respect to $\half{\rm p}^2$ with a relative bound
$\kappa_j\geq 0$;
\item[(3)]
$\d A\lk1-({\kappa_1+\kappa_2+\kappa_3})\rk^{-\han}<1$;
\item[(4)]
$\d \int_{\RR^3} \frac{|\log(\half\sqrt{b_1(y)^2+b_2(y)^2})|} {2\pi  |x-y|} dy <\infty,\quad \mbox{a.e.} \;\; x\in\RR^3$.
\end{enumerate}
Then the relativistic Schr\"odinger operator $h_\han ^{\rm rel}+V$ (resp. $h_{\mathbb Z_2}^{\rm rel}+V$) is essentially
self-adjoint on $\mathbb C^2\otimes C_0^\infty(\RR^3)$ (resp. $\ell^2(\mathbb Z_2)\otimes C_0^\infty(\RR^3)$) and
\eq{ko16}
(f, e^{-t(h_{\mathbb Z_2}^{\rm rel} + V)}g)=\sum_{\alpha=1,2}\int_{\RR^3} dx
\mathbb E_{P\times \mu\times \nu}^{x,\alpha,0}\lkkk e^{T_t^\bern}
\ov{f(B_0,\c_{N_0})} g(B_{T_t^\bern}, \c_{N_{T_t^\bern}})e^{\SSS^\bern} \rkkk,
\en
where
$\c_{N_{T_t^\bern}}=(-1)^{N_{T_t^\bern}}$, the subordinator $T_t^\bern$ is defined by $T_t^\bern=\inf\{s>0 \,|\,
B_s+m s = t\}$ and the exponent $\SSS^\bern=\SSS_V^\bern +\SSS_a^\bern +\SSS_{\rm spin}^\bern$ is given by
\begin{eqnarray*}
&&
\SSS_V^\bern  =-\int_0^t V(B_{T_s^\bern}) ds, \\
&&
\SSS_a^\bern  =-i\int_0^{T_t^\bern } a(B_s)\circ dB_s,\\
&&
\SSS_{\rm spin}^\bern = \int_0^{T_t^\bern } \half b_3(B_s)\c_{N_s} ds + \int_0^{T_t^\bern+} \log\lk
\half \lk b_1(B_s)-i\c _{N_{s-}}b_2(B_s)\rk\rk dN_s.
\end{eqnarray*}
\et
\proof
Set $S=-\half \mmm {b_3}{\sqrt{b_1^2+b_2^2}}{\sqrt{b_1^2+b_2^2}}{-b_3}$. We see that $S$ is relatively bounded with
respect to $\half {\rm p}^2\mmm 1 0 0 1$ with a relative bound $\kappa=\kappa_1+\kappa_2+\kappa_3$.
Note that
$$
\|\sqrt{{\rm p}^2+m^2}f\|^2=(f, ({\rm p}^2+m^2)f)=\|(h_\han^{\rm rel}(0)+m)f\|^2 +
 2(f, -Sf+\is{h_\han^0}f).
$$
Since $|(f, S f)|\leq \kappa(f, \half{\rm p}^2 f)+\kappa' \|f\|^2$, with a constant $\kappa'$, we have
$$
\|\sqrt{{\rm p}^2+m^2}f\|^2 \leq \|
(h_\han^{\rm rel}(0)+m)f\|^2+{\kappa}
\|\sqrt{{\rm p}^2+m^2}f\|^2
+(|\is{h_\han^0}|+\kappa')\|f\|.
$$
Together with $\|Vf\|\leq A\|\sqrt{{\rm p}^2+m^2}f\|+A'\|f\|^2$,
with a constant $A'$, we have
$$
\|Vf\| \leq A\lk 1-{\kappa}\rk ^{-\han}\|h_\han^{\rm rel}(0)f\|+
\lk A'+Am+A\sqrt{2|\is{h_\han^0}|+\kappa'}\rk\|f\|.
$$
Thus by assumption (3) above, $V$ is relatively bounded with respect to $h_\han^{\rm rel}(0)$ with relative bound
$A\lk 1-\kappa\rk ^{-\han}<1$, and hence essential self-adjointness of $h_\han^{\rm rel}+V$ on $\mathbb
C^2 \otimes C_0^\infty(\RR^3)$ follows by Theorem \ref{ko39}. Since
$$
\int_0^\infty ds\int_{\RR^3} dy (2\pi s)^{-3/2} e^{-\frac{|x-y|^2}{2s}} |\log(U_1(y))| =
\int_{\RR^3} \frac{|\log(\half \sqrt{b_1(y)^2+b_2(y)^2})|} {2\pi  |x-y|} dy <\infty,
$$
\kak{cont22} or \kak{cont23} is satisfied. Then \kak{ko16} follows from Theorem \ref{MAIN}.
\qed

We further have the energy comparison inequality following by \kak{diarel}.
Let
\begin{eqnarray*}
&&
\tilde h_\han^{\rm rel}=\sqrt{2(h_\han-\is{h_\han^0})+m^2}-m,\\
&&
\tilde h_{\mathbb Z_2} ^{\rm rel}=\sqrt{2(h_{\mathbb Z_2}-\is{h_{\mathbb Z_2}^0})+m^2}-m.
\end{eqnarray*}
Note that $h_\han-\is{h_\han^0}\geq 0$ and $h_{\mathbb Z_2}-\is{h_{\mathbb Z_2}^0}\geq0$ by \kak{bern20}
in Corollary \ref{wr4}.
\bc{dia4}{\bf (Diamagnetic inequality)}
Under the assumptions of Theorem \ref{ko50}
\eq{diarel}
\left|\lk f, e^{-t(\tilde h_{\mathbb Z_2} ^{\rm rel} +V)}g\rk\right| \leq
\lk |f|, e^{-t(h_{\mathbb Z_2} ^{\rm rel}(0) +V)}|g|\rk.
\en
In particular, it follows that
$$
\is{h_{\mathbb Z_2} ^{\rm rel}(0)+V} \leq \is{\tilde h_{\mathbb Z_2} ^{\rm rel}+V},
$$
or equivalently
$$
\is{h_\han^{\rm rel}(0)+V} \leq \is{\tilde h_\han ^{\rm rel}+V}.
$$
\ec

\bigskip
\noindent
\emph{Generator of $\xi_{T_t^\bern}$. \quad}
In \cite{hl08} and in Lemma \ref{Main} above we used the $\RR^3\times\mathbb Z_2$-valued joint
Brownian and jump process $\xi_t=(B_t,\c_{N_t})$ starting from $\xi_0=(x,\c_\alpha)$ to get the
functional integral representation for Schr\"odinger operators with spin $\han$. The generator of
this process is $(\han){\rm p}^2+\s_{\rm F}$, where $\s_{\rm F}$ is the fermionic harmonic
oscillator defined in terms of the Pauli matrices by
$$\s_{\rm F}=(\han) (\s_3+i\s_2)(\s_3-i\s_2)-
(\han) I=-\s_1.$$
Note that $\is {\half {\rm p}^2+\s_F}=-1$. Similarly, we can identify
the generator for the subordinated joint Brownian and jump process
$$
\xi_{T_t^\bern} = (B_{T_t^\bern}, \c_{N_{T_t^\bern}}), \quad \BM \times \PO \times\SU
\rightarrow \RR^3\times \mathbb Z_2
$$
starting at $\xi_0=(x,\c_\alpha)$ to be
$$
G=\bern\lk \half{\rm p}^2+\s_{\rm F}+1\rk.
$$
This is obtained from the relationship
$$
\sum_{\alpha=1,2} \int \EE_{P\times\mu\times\nu}^{x,\alpha,0} \left[e^{-T_t^\bern}
\ov{f(\xi_0)}g(\xi_{T_t^\bern}) \right] dx =(f, e^{-tG}g)
$$
under the identification $L^2(\RR^3;\CC^2)\cong L^2(\RR^3\times\mathbb Z_2)$.

\bigskip
\noindent
\emph{Support of magnetic field. \quad}
Consider the case when $b_1(x)-i \c  b_2(x)$ vanishes for some $x\in \BR$. In this case it is not clear
whether $\int_0^{t+}|\log \half(b_1(B_s)- i\c _{N_{s-}} b_2(B_s)) |dN_s$ is almost surely finite and
assumption (4) in Theorem \ref{ko50} holds at all. An example when this is not the case is obtained by
choosing $b\in (C_0^\infty(\RR^3))^3$. To improve Theorem \ref{ko50} we use the ideas of \cite{hl08},
where we considered this problem for the Schr\"odinger operator $\half(\VS \cdot ({\rm p}-a))^2+V$. Let
$\delta_\epsilon(z)= \lkk
\begin{array}{ll} 1,&|z|<\epsilon/2,\\
0,&|z|\geq \epsilon/2,
\end{array}
\right.
$for $z\in\CC$ and set $\chi_\epsilon (z)=z+\epsilon \delta_\epsilon(z)$. We see that
$$
\left|\chi_\epsilon \lk -\half (b_1(x)-i\c  b_2(x)) \rk\right| >\epsilon/2,\quad
(x,\c)\in \RR^3\times\mathbb Z_2.
$$
Define $h_{\mathbb Z_2} ^\epsilon$ by $h_{\mathbb Z_2} $ with the off-diagonal part replaced by
$\chi_\epsilon\lk \skima-\half(b_1(x)-i\c b_2(x)) \rk$, i.e.,
$$
h_{\mathbb Z_2} ^\epsilon f(x,\c) = \lk h-\half \c  b_3(x)\rk f(x,\c )+\chi_\epsilon \lk
-\half (b_1(x)-i\c  b_2(x)) \rk f(x,-\c).
$$
We also see that $h_{\mathbb Z_2} ^\epsilon$ is self-adjoint on $D(h)$. Define $h_{\mathbb Z_2}^
{{\rm rel},\epsilon} = \sqrt{2 \ov{h_{\mathbb Z_2} ^\epsilon}+m^2}-m$, where $\ov{h_{\mathbb Z_2}
^\epsilon} = {h_{\mathbb Z_2} ^\epsilon} - \is{{h_{\mathbb Z_2} ^\epsilon}}$ as usual. Since
${h_{\mathbb Z_2} ^\epsilon}$  converges to ${h_{\mathbb Z_2}} $ as $\epsilon\downarrow 0$ in
uniform resolvent sense, $\is{{h_{\mathbb Z_2} ^\epsilon}} \to \is{{h_{\mathbb Z_2} }}$ as
$\epsilon\downarrow 0$. Under the assumptions of Theorem \ref{ko50} but without assuming (4) there
we are able to show that $h_{\mathbb Z_2}^{{\rm rel},\epsilon}$ is essentially self-adjoint on
$\ell^2(\mathbb Z_2)\otimes C_0^\infty(\RR^3)$ and the functional integral representation of
$h_{\mathbb Z_2} ^{{\rm rel},\epsilon} +V$ holds by \kak{ko16} with
\begin{eqnarray*}
&&
\SSS_{\rm spin}^\bern (\epsilon) = \int_0^{T_t^\bern} \lk
\half b_3(B_s) \c_{N_s}-\is{h_{\mathbb Z_2}^\epsilon}\rk ds\\
&&
\hspace{4cm}
+\int_0^{T_t^\bern+} \log \lk -\chi_\epsilon \lk -\half (b_1(B_s)-i \c_{N_{s-}} b_2(B_s)) \rk \rk dN_s
\end{eqnarray*}
instead of $\SSS_{\rm spin}^\bern$. Moreover, $h_{\mathbb Z_2} ^{{\rm rel},\epsilon}+V$ converges to
$h_{\mathbb Z_2} ^{{\rm rel}} +V$ on the common core $\ell^2(\mathbb Z_2)\otimes C_0^\infty(\RR^3)$
so that
$$
\lim_{\epsilon\downarrow 0} \exp\left(-t (h_{\mathbb Z_2} ^{{\rm rel},\epsilon}+V)\right) =
\exp\left(-t(h_{\mathbb Z_2}^{{\rm rel}}+V)\right)
$$
in strong sense. Hence we have the theorem below.
\bt{ko51}
Take Assumption (A4) and assumptions (1)-(3) in Theorem \ref{ko50}. Then the functional integral
representation for $h_{\mathbb Z_2} ^{\rm rel}+V$ is given by
\eq{ko16new}
(f, e^{-t(h _\han^{\rm rel}+V)}g) = \lim_{\epsilon\downarrow 0} \sum_{\alpha=1,2}
\int_{\RR^3} dx \mathbb E_{P\times \mu\times \nu}^{x,\alpha,0}\lkkk e^{T_t^\bern} \ov{f(B_0,\c _{N_0})}
g(B_{T_t^\bern}, \c _{N_{T_t^\bern}})e^{\SSS^\bern (\epsilon)} \rkkk,
\en
where $\SSS^\bern (\epsilon)=\SSS_V^\bern +\SSS_a^\bern + \SSS_{\rm spin}^\bern (\epsilon)$.
\et

\subsection{Spinless case}
Finally consider the spinless case and write
\begin{eqnarray}
&&
h^{\rm rel}=\sqrt{({\rm p}-a)^2+m^2}-m,\\
 &&
h^{\rm rel}(0)=\sqrt{{\rm p}^2+ m^2}-m.
\end{eqnarray}
\bt{ko52}
Let Assumption (A3) hold and $V$ be relatively bounded with respect to $\sqrt{{\rm p}^2+m^2}$ with
relative bound strictly less than 1. Then $h^{\rm rel}+V$ is essentially self-adjoint on
$C_0^\infty(\RR^3)$ and
\eq{ko42}
(f, e^{-t(h^{\rm rel}+V)}g)=\int_{\RR^3} dx \mathbb E_{P\times  \nu}^{x,0} \lkkk \ov{f(B_0)}
g(B_{T_t^\bern}) e^{\SSS_V^\bern +\SSS_A^\bern } \rkkk.
\en
\et
\proof
The essential self-adjointness follows from (2) of Corollary \ref{ko21}, and \kak{ko42} from
Theorem \ref{f}.
\qed

By Theorem \ref{ko52}  we also have the following energy comparison inequality.
\bc{dia3}{\bf (Diamagnetic inequality)}
Under the assumptions of Theorem \ref{ko52}
\eq{diarel2}
|(f, e^{-t(h^{\rm rel}+V)}g)| \leq (|f|, e^{-t(h^{\rm rel}(0)+V)}|g|)
\en
and
$$
\is {h^{\rm rel}(0)+V}\leq \is {h^{\rm rel}+V}.
$$
\ec
In the case of $\bern(u)=\sqrt{2u+m^2}-m$, Assumptions \ref{a} and \ref{b} are readily satisfied. Furthermore,
by Theorem \ref{hyp} we have the result below.
\bc{hyp3}
{\bf (Hypercontractivity)}
Let the assumptions of Theorem \ref{ko52} and one of the three equivalent conditions in Proposition
\ref{d} with $\bern(u)=\sqrt{2u+m^2}-m$ hold. Then $e^{-t(h^{\rm rel}+V)}$ is a bounded operator from
$L^p(\BR)$ to $L^q(\BR)$ for all $1\leq p\leq q\leq\infty$.
\ec

\appendix{}
\section{Appendix}
For a given L\'evy process $(L_t)_{t\geq 0}$ on a probability space $(\Omega, \ms F, P)$ the notation
$dL_t=L_t-L_0$ is used for its differential. Let $F\in C^2(\RR)$. The differential of the transformed
process $dF(L_t)$ can be computed by the following It\^o formula.

\begin{proposition}{\rm (\textbf{It\^o formula})}
Let $\ms F_t$ be the natural filtration $\s((B_s, N_s^\beta), 0\leq s\leq t,\beta=1,...,p)$. Consider
$$
L_t^i =\int_0^t  f^i(s,\omega)  ds +
\int_0^t  g^i(s,\omega)  \cdot dB_s+ \sum_{\beta=1}^{p-1} \int_0^{t+}
h_\beta^i (s,\omega) dN_s^\beta, \quad i=1,...,n
$$
where $f^i(\cdot,\omega)\in L_{\rm loc}^1(\RR)$ a.s, $g^i\in \ms E_{\rm loc}$ and $h_\beta^i(s,\omega)$
is adapted with respect to $\ms F _t$, left continuous in $s$ and $\int_0^{t+}|h_\beta^i(s,\omega)|
dN_s^\beta<\infty$ a.s.
Take $F\in C^2(\RR^n)$. Then for the random process $F(L_t)$ the expression
\begin{eqnarray*}
dF(L_t)
&=&
\sum_{i=1}^n \int_0^t F_i(L_s) f^i(s)ds+\sum_{i,j=1}^n \int_0^t \half F_{ij}(L_s) g^i(s)\cdot g^j(s)ds \\
&&
+ \sum_{i=1}^n \int_0^t F_i(L_s) g^i(s)\cdot  dB_s + \sum_{\beta=1}^{p-1} \int_0^{t+}
(F(L_{s-}+h_\beta(s))-F(L_{s-})) dN_s^\beta
\end{eqnarray*}
holds. Here $F_i=\partial_i F$ and $F_{ij}=\partial_i\partial_j F$.
\end{proposition}

Furthermore, the following form of the product rule holds.
\begin{proposition}{\rm (\textbf{Product\ rule})}
Let $(L_t)_{t\geq 0}$ and $(M_t)_{t\geq 0}$ be two random processes. Then $d(L_t M_t)=
dL_t\cdot M_t+L_t\cdot dM_t+dL_t\cdot dM_t$, computed by the rules $dt dt=0$, $dB_t^\mu dt=0$,
$dB_t^\mu dB_t^\nu=\delta_{\mu\nu} dt$, $dN_t^\alpha dN_t^\beta=0$, $dN_t^\alpha dt=0$, and
$dN_t^\alpha dB_t=0$.
\end{proposition}
For proofs see, for instance, \cite{iw, LHB09}.

\bigskip
\noindent {\bf Acknowledgments:}
FH acknowledges support of Grant-in-Aid for Scientific Research (B) 20340032 from JSPS and is thankful
to Loughborough University, Paris XI University, and IHES, Bures-sur-Yvette, for hospitality. TI
acknowledges support of JSPS Grant-in-Aid for Scientific Research (C) 20540161. JL thanks Royal Society
for an international travel grant, and the hospitality of Erwin Schr\"odinger Institute, Vienna, IHES,
Bures-sur-Yvette, and Kyushu University, where various parts of this work have been done.

\end{document}